\documentclass{aa}

\usepackage{graphicx,epstopdf}
\usepackage{amsmath,mathtools}
\usepackage{amssymb}
\usepackage{txfonts}

\def\kmps{km\,s$^{-1}$}

\def\mrats{g\,cm$^{-2}$\,s$^{-1}$}

\def\mrat{M$_\odot$\,yr$^{-1}$}

\def\fint{ergs\,cm$^{-2}$\,s$^{-1}$}
\def\flux{ergs\,cm$^{-2}$\,s$^{-1}$\,\AA$^{-1}$}
\def\crat{s$^{-1}$}
\def\cps{s$^{-1}$}
\def\msun{M$_\odot$}

\def\amh{AM Herculis}
\def\xmmn{{\it XMM-Newton}}
\DeclarePairedDelimiter\abs{\lvert}{\rvert}
\begin{document} 

\title{The various accretion modes of AM Herculis: Clues from 
multi-wavelength observations in high accretion states\thanks{Based on
  observations obtained with \xmmn, an ESA science  mission with instruments  and contributions directly funded by ESA member states and NASA. Based on data obtained with the STELLA robotic telescopes in Tenerife, an AIP (Leibniz-Institut für Astrophysik Potsdam) facility jointly operated by AIP and IAC (Instituto de Astrofísica de Canarias).
Based on observations collected at the Centro Astronómico Hispano Alemán (CAHA) at Calar Alto, operated jointly by the Max-Planck Institut für Astronomie and the Instituto de Astrofísica de Andalucía (CSIC). Based also on the effort of 35 amateurs distributed worldwide, and organized by the American Association of Variable Star Observers (AAVSO) who contributed photometric observations.}
} 
\author{
A.D. Schwope\inst{1}
\and
H. Worpel\inst{1}
\and
I. Traulsen\inst{1}
\and 
D. Sablowski\inst{1}
}
\titlerunning{The various accretion modes of AM Herculis}
\institute{
Leibniz-Institut f\"ur Astrophysik Potsdam (AIP),
An der Sternwarte 16, 14482 Potsdam, Germany}

\date{Received February 12, 2020; accepted August 6, 2020 }
 
\abstract{We report on XMM-Newton and NuSTAR X-ray observations of the prototypical polar, AM Herculis, supported by ground-based photometry and spectroscopy, all obtained in high accretion states. In 2005, AM Herculis was in its regular mode of accretion, showing a self-eclipse of the main accreting pole. X-ray emission during the self-eclipse was assigned to a second pole through its soft X-ray emission and not to scattering. In 2015, AM Herculis was in its reversed mode with strong soft blobby accretion at the far accretion region. The blobby acretion region was more luminous than the other, persistently accreting, therefore called main region. Hard X-rays from the main region did not show a self-eclipse indicating a pronounced migration of the accretion footpoint. Extended phases of soft X-ray extinction through absorption in interbinary matter were observed for the first time in AM Herculis. The spectral parameters of a large number of individual soft flares could be derived. Simultaneous NuSTAR observations in the reversed mode of accretion revealed clear evidence for Compton reflection of radiation from the main pole at the white dwarf surface. This picture is supported by the trace of the Fe resonance line at 6.4 keV through the whole orbit. Highly ionized oxygen lines observed with the Reflection Grating Spectrometer (RGS) were tentatively located at the bottom of the accretion column, although the implied densities are quite different from expectations. In the regular mode of accretion, the phase-dependent modulations in the ultraviolet (UV) are explained with projection effects of an accretion-heated spot at the prime pole. In the reversed mode projection effects cannot be recognized. The light curves reveal an extra source of UV radiation and extended UV absorbing dips. An H$\alpha$ Doppler map obtained contemporaneously with the NuSTAR and XMM-Newton observations in 2015 lacks the typical narrow emission line from the donor star but reveals emission from an accretion curtain in all velocity quadrants, indicating widely dispersed matter in the magnetosphere.}


\maketitle
%

\section{Introduction}
\amh\ is the prototype of its class of magnetic cataclysmic binaries, also 
dubbed the polars \citep{cropper90}. A polar consists of an accreting strongly 
magnetic white dwarf (typically labeled as the primary) and a late-type
secondary star, the donor, that loses matter through the inner Lagrangian 
point. The magnetic field of the white dwarf keeps both stars in synchronous rotation and prevents the formation of an accretion disk, but guides the accreted matter toward the polar accretion regions where it is instrumental in releasing the gravitational energy in the form of cyclotron radiation in the infrared, optical, and ultraviolet spectral ranges.

\amh\ was among the brightest X-ray sources in the sky and was already among the 
125 sources reported in the second Uhuru catalog \citep{giacconi+72}. 
A composite spectrum with a low- and a high-temperature component (or a steep
and a flat spectral component) was already found from OSO-8 observations
performed in 1975 \citep{bunner78} and ultraviolet observations with the International Ultraviolet Explorer (IUE) appeared to confirm
the blackbody interpretation for the soft X-ray flux \citep{raymond+79}.
An investigation based on HEAO 1 observations accumulated in 1978 revealed
evidence for a white-dwarf reflected X-ray component and unrelated soft and 
hard X-ray spectral components \citep{rothschild+81}, later confirmed by a
timing analysis of {\it EINSTEIN} observations of the source \citep{stella+86}.

\begin{figure*}[th]
\resizebox{\hsize}{!}{\includegraphics[angle=-90,clip=]{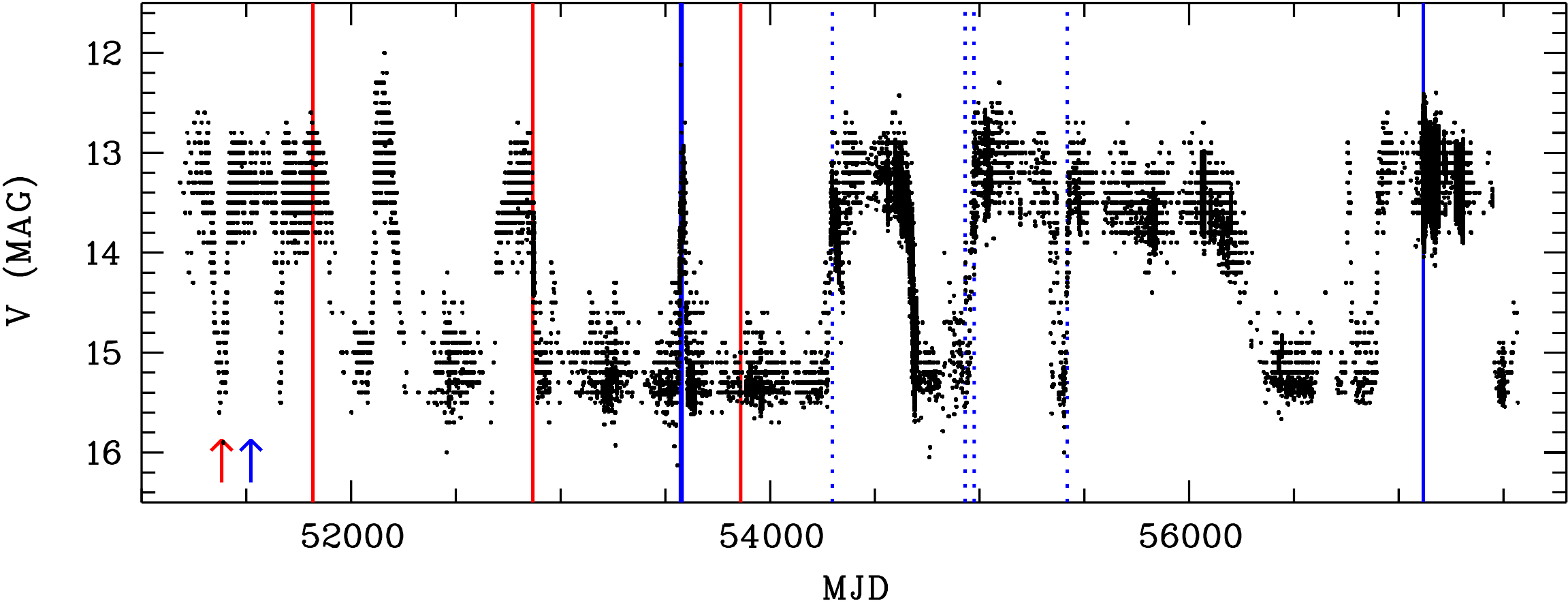}}
\caption{AAVSO V-band light curve between January 1, 1999, and June 20, 2016.
Launch dates of \xmmn\ and Chandra are indicated by blue and red arrows. X-ray
observations with those observatories are indicated by vertical lines. \xmmn\ slew observations are indicated with dotted lines.
\label{f:aavso_99_16}}
\end{figure*}

Observations prior to 1983 were mostly interpreted in the framework of one accreting pole located roughly opposite to the donor star, which was self-eclipsed by the white dwarf during the revolution of the 3.09 h binary. Complexities in the polarization and X-ray light curves led to speculations about a second accretion region at the donor star-facing hemisphere \citep{kruszewski78,szkody+80,tapia77}. A two-pole scenario was confirmed by EXOSAT observations performed in 1983, when bright soft X-rays,  modulated on the white-dwarf rotation period, were seen during the hard X-ray eclipses \citep{heise+85}. Optical light curves obtained in this reversed mode of accretion were found to be surprisingly similar to those obtained in the normal mode \citep{mazeh+86}. The occurrence of the reversed mode of accretion led to the further development of the blobby accretion scenario originally described by \citet{kuijpers_pringle82} to account for the shape of the soft X-ray light curves and the energy balance in that particular mode of accretion \citep{hameury_king88, litchfield_king90}.

Being the prototype and brightest member of its class, \amh\ has been observed by almost every X-ray observatory. Various aspects of accretion physics were addressed using these data. Of particular interest were the origin of the various emission lines, the
occurrence of absorption edges at soft energies, the influence of absorbers 
and reflectors on the spectral shape, and the 
soft versus hard emission from the accretion column. Accounts on the energy
balance between soft and hard X-ray emission using HEAO-1, ROSAT, ASCA, EUVE, RXTE, and Chandra were given by
\citep{rothschild+81,ramsay+96,ishida+97, christian00, beuermann+08}, and \cite{beuermann+12}, respectively.
The first X-ray CCD spectrum obtained with ASCA revealed the presence of a
multi-temperature thermal plasma spectrum with a superposed Fe-line structure between 6.4 and 7.0 keV \citep{ishida+97}. More recent Chandra observations with both gratings (LETG, HETG), were intended to obtain a soft X-ray spectral model for the normal mode of accretion and to study possible radial velocity variations of emission lines in the energy range covered by the HETG \citep{girish+07, beuermann+12}.

\amh\ was in the blind spot of the \xmmn\ X-ray sky at the beginning of
the mission and thus did not become part of any guaranteed time program. The
geometrical configuration changed slowly and exploratory observations were
possible in 2005 \citep{schwope+06}. 
These observations were luckily obtained in a brief high accretion
state but with non-optimum camera settings. A decent \xmmn\ observation
was still outstanding and hampered by extended periods of low activity of the
source. \amh\ was thus made the target of a target-of-opportunity proposal to
be triggered by an optical monitoring program indicating a stable high
accretion state. This happened during the spring observation window in
2015 and was indicated by sporadic amateur observations, the results of which were reported via the AAVSO web pages\footnote{http://www.aavso.org}. 
The high state was confirmed through 
time-resolved photometry with the robotic 
telescope Stella-I located on the island of Tenerife. Following this, a multi-observatory campaign was launched involving \xmmn, NuSTAR, Calar Alto, and the AAVSO.

The results of both campaigns, in 2005 and 2015, are reported in this paper.
We always use the spectroscopic ephemeris 
\begin{equation}
T_0(HJD) = 2451763.452523 + E \times 0.128927103
\end{equation} 
to relate time and binary orbital phase. Phase zero at $T_0$
refers to the inferior conjunction of
the secondary star as determined by \citet{schwarz+02}. The phase difference between our convention and the otherwise used magnetic phase is $\phi_{\rm orb} = \phi_{\rm mag} + 0.367$.

Satellite data used in the current paper 
are given in TT (Terrestrial Time; which is the same as Barycentric Dynamical Time TDB in the context (hence accuracy)  of this paper), time zero refers to TDB =  2451763.453266. The difference between heliocentric and barycentric times was about 0.5\,s at August 6, 2000, and is of no relevance for the current paper.

\begin{table*}
\caption{\xmmn\ and NuSTAR observations of \amh\ in 2005 and 2015. Given are
  dates and times for the start of the observations, the observation ID,
  information about the camera modi and filters used as well as the
  exposure times. The latter is given in seconds and always somewhat rounded.
PN stands for EPIC-pn, MOS for EPIC-MOS, TI for timing mode,
SW for small window mode, PW2/3 for PrimePartialWindow2/3, TCK for the thick, and
THN for the thin filter. For details see the \xmmn\ users handbook.}
\begin{tabular}{lrrlr}
Start Obs& OBSID & revolution & camera-mode-filter & exposure\\
\hline
\multicolumn{5}{l}{\underline{\xmmn}}\\[0.5ex]
2005-07-19 16:48:42 & 0305240201 & 1027 & PN-TI-THK / MOS-PW3-THN & 8240/9350\\
& & & OM-TI-UVW2 & $2 \times 4307$ \\
2005-07-21 16:45:28 & 0305240101 & 1028 & PN-TI-THK / MOS-PW3-THN & 8220/8880\\
& & & OM-TI-UVW2 & $1 \times 4307$ \\
2005-07-23 16:34:00 & 0305240301 & 1029 & OM-TI-UVW2 & $2 \times 4307$ \\
2005-07-25 16:25:03 & 0305240401 & 1030 & PN-TI-THK / MOS-PW3-THN & 8240/9270\\
& & & OM-TI-UVW2 & $2 \times 4307$ \\
2005-07-27 16:17:00 & 0305240501 & 1031 & PN-TI-THK / MOS-PW3-THN & 8240/9370\\
& & & OM-TI-UVW2 & $2 \times 4307$ \\
2015-04-06 02:36:01 & 0744180801  & 2806 & PN-SW-THK / MOS-PW2-THK & 30450,42300\\
& & & OM-TI-UVM2 & $7 \times 3540$ \\
& & & OM-UV-grism & $2 \times 3200, 2\times 3600$\\[1ex]
\multicolumn{5}{l}{\underline{NuSTAR}}\\[0.5ex]
2015-04-05 13:11:07 & 80001037001 & & & 56000 \\
\hline
\end{tabular}
\label{t:obs}
\end{table*}

\begin{figure}
\resizebox{\hsize}{!}{\includegraphics[clip=]{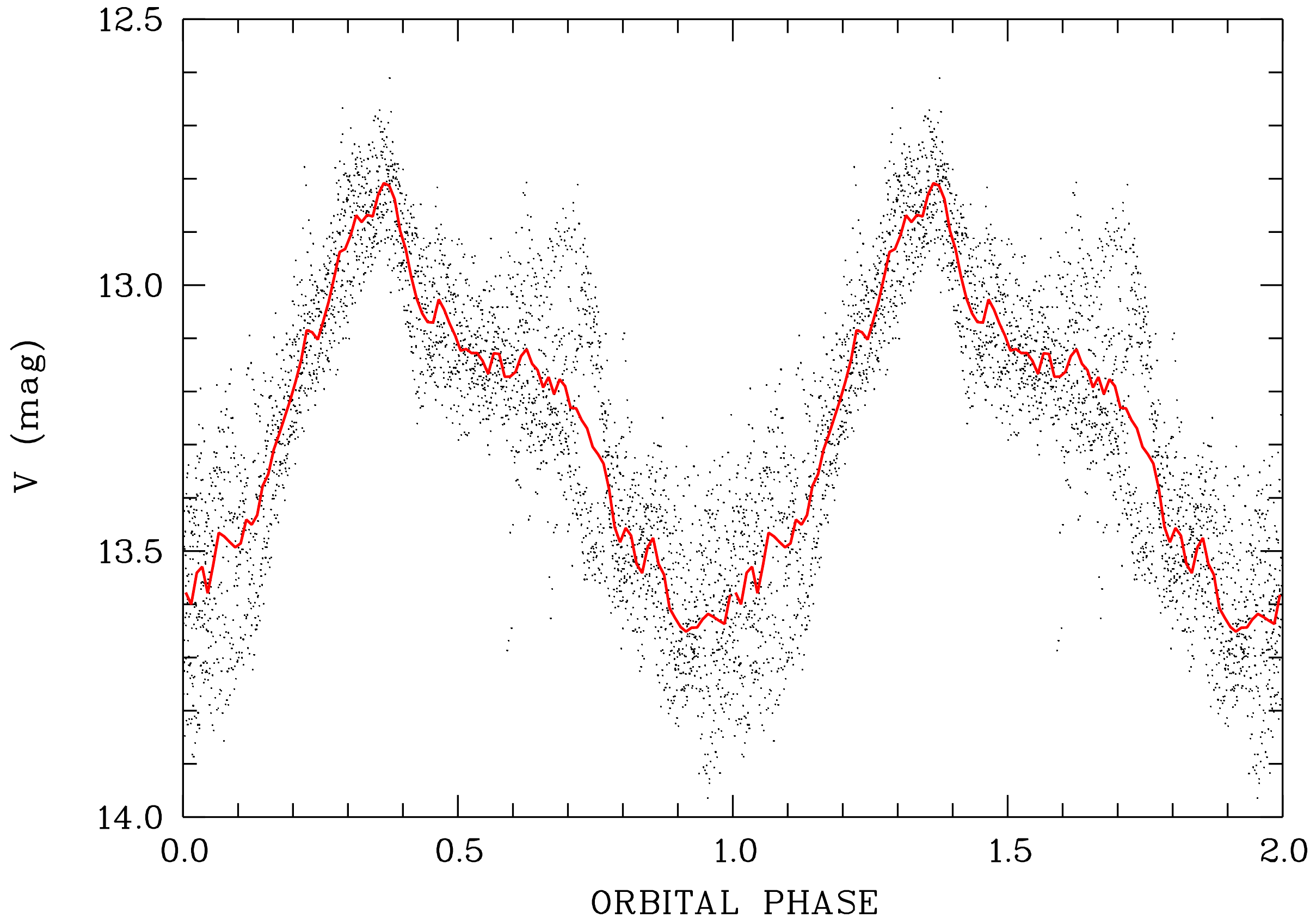}}
\caption{V-band photometry obtained with STELLA/WiFSIP between March 27 and April  8, 2015. Shown is the average behavior binned into 100 phase bins (red
  line) on the background of the 3200 individual photometric data points
  illustrating the variability of the source. The typical photometric error is
0.005 mag.
\label{f:stella}}
\end{figure}

\begin{figure}
\resizebox{\hsize}{!}{\includegraphics[clip=]{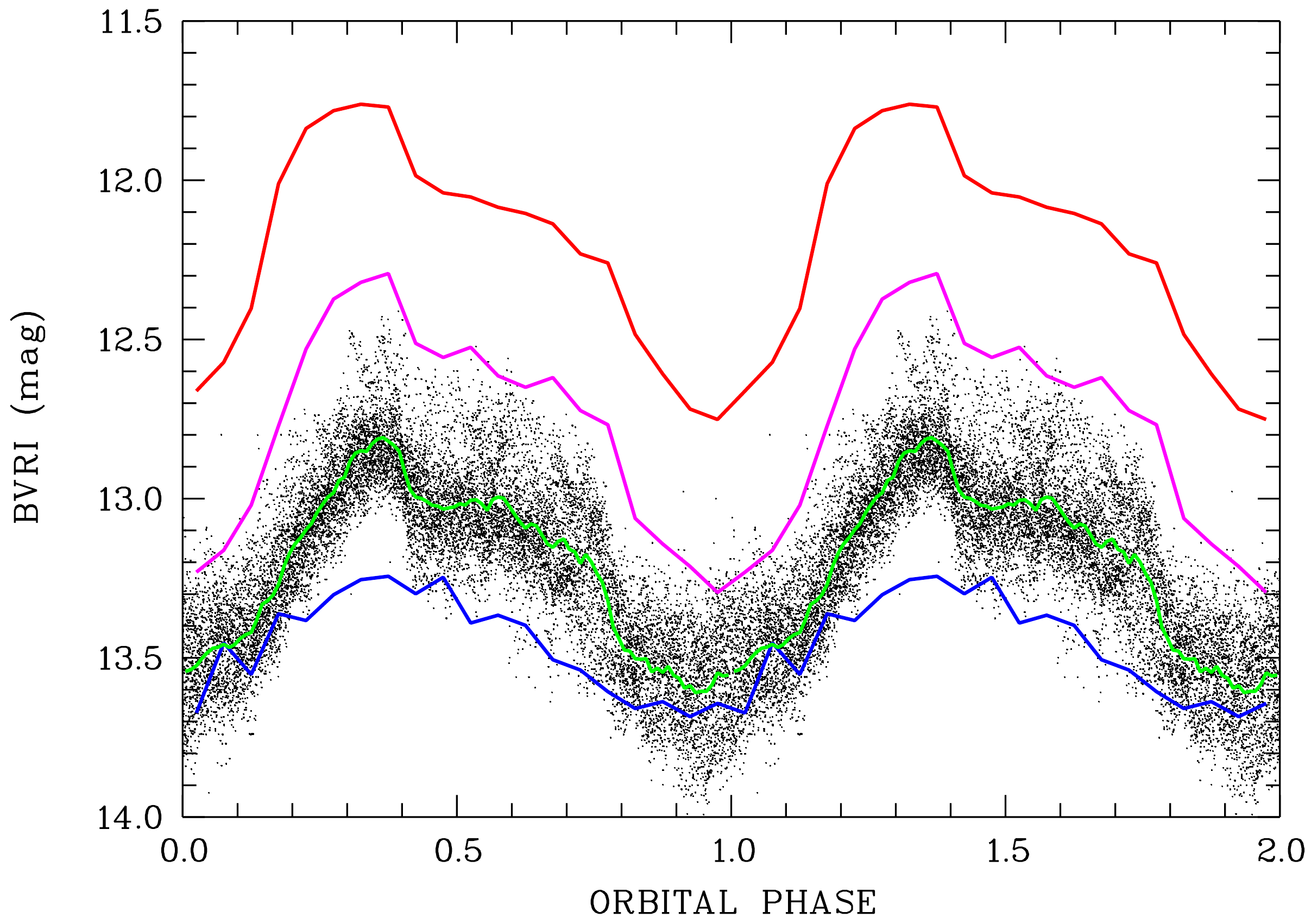}} 
\caption{$BVRI$ photometry of \amh\ collected between April 2 and April 16 by amateur astronomers and provided via the AAVSO web-pages. The 15037 individual measurements that were taken through a $V$ filter are shown with small black symbols. There are 272/202/202 measurements through $B,R,$ and $I$ filters, respectively, that are not shown individually. Bin sizes for phase-folding the original data were 0.01 for $V$ (green line) and 0.05 otherwise, which are shown in blue/magenta/red for the $BRI$ filters.
\label{f:aavsofol}}
\end{figure}

\section{Observations and analysis}\label{s:obs}
\subsection{Satellite X-ray and ultraviolet observations with \xmmn\ and
  NuSTAR} 

\amh\ was the target of two observation campaigns with \xmmn, the first in 2005, the second in 2015. Observations in 2015 were performed simultaneously with NuSTAR to
complement the energy range of \xmmn\ toward higher energies.

In 2005 during \xmmn\ (AO4) the whole observation that was supposed to last 36
ks was broken into chunks of about 10\,ks. Originally three visits of the
source were planned but given the limited visibility of the target at the end
of the spacecraft revolution and the potential loss of data due to high
background radiation, a total of five visits were generously scheduled by the 
\xmmn\ SOC. These were performed in five consecutive spacecraft orbits of which all five yielded data from the OM and of which four yielded X-ray data that are accessible through the \xmmn\ Science Archive (XSA) and were used in our analysis. 

Details on the amount of data and the camera settings of the 
high-energy satellite observations are given in Table~\ref{t:obs}. The table
lists all instruments except the RGS \citep{denHerderEtAl2001}, which was 
operated during all observations
in standard mode. The data from \xmmn\ revolution 1029 were affected by high
background radiation and revealed no X-ray data.

In 2015 coordinated \xmmn and NuSTAR observations were triggered through an optical
monitoring program that was indicating a stable high accretion state. This
time a long uninterrupted observation with \xmmn\ could be achieved, thus
covering 2.7/3.8 orbital cycles of the binary with EPIC-pn/MOS \citep{StruderEtAl2001, TurnerEtAl2001}. 

Countrate estimates prior to the observations demanded special care to avoid
pile-up, for example during phases of extremely soft blobby accretion. 
Two different solutions were attempted for the two observations. In 2005 
EPIC-pn was used in timing mode (killing all soft photons below $<$0.4 keV) 
with the thin filter and in 2015 in small window mode with the thick filter (killing soft photons by reduced efficiency). EPIC-MOS was used at both occasions with one of the imaging modes. It was regarded being the optimized detector for a low state of AM Herculis.

The optical monitor (OM, \citealt{MasonEtAl2001}) onboard \xmmn\ was mostly used in timing mode to
generate orbital phase-resolved light curves. Observations through the UVW2 
filter in 2005
revealed a maximum count rate of about 20 \crat. It was replaced in 2015 by
UVM2 with twice the effective area and longer effective wavelength (231\,nm 
instead of 212\,nm) revealing up to 60 \crat. 
In 2015 also four UV-grism spectra were obtained with
integration times between 3200 and 3600\,s.

NuSTAR is in a low Earth orbit and continuous observations of a target are not
possible \citep{harrison+13}. Observations are therefore broken up into smaller
chunks. The net granted observation time with NuSTAR was longer than that with \xmmn. 
NuSTAR observations began before the \xmmn\ observations and were finished afterward.
NuSTAR observations thus were spread over 29.5 hours and divided into 19 chunks.
Both telescopes of the spacecraft delivered science-ready data.

\begin{table}
\caption{STELLA photometry of \amh\ in March and April 2015. All exposures were 
obtained through a Bessel V-filter with exposure time 15 seconds.\label{t:stella}}
\begin{tabular}{llr}
Date Start & Start/End [UT]& \# frames \\
\hline
2015-03-27 & 02:10 -- 05:43 & 443 \\
2015-03-31 & 02:39 -- 06:12 & 443 \\
2015-04-04 & 00:01 -- 06:29 & 774 \\ 
2015-04-05 & 00:10 -- 06:28 & 754 \\
2015-04-07 & 23:49 -- 06:07 & 788 \\
\hline
SUM & 26.2 hours & 3202 \\
\end{tabular}
\end{table}

\begin{figure*}
\resizebox{1.0\hsize}{!}{\includegraphics[angle=-90,clip=]{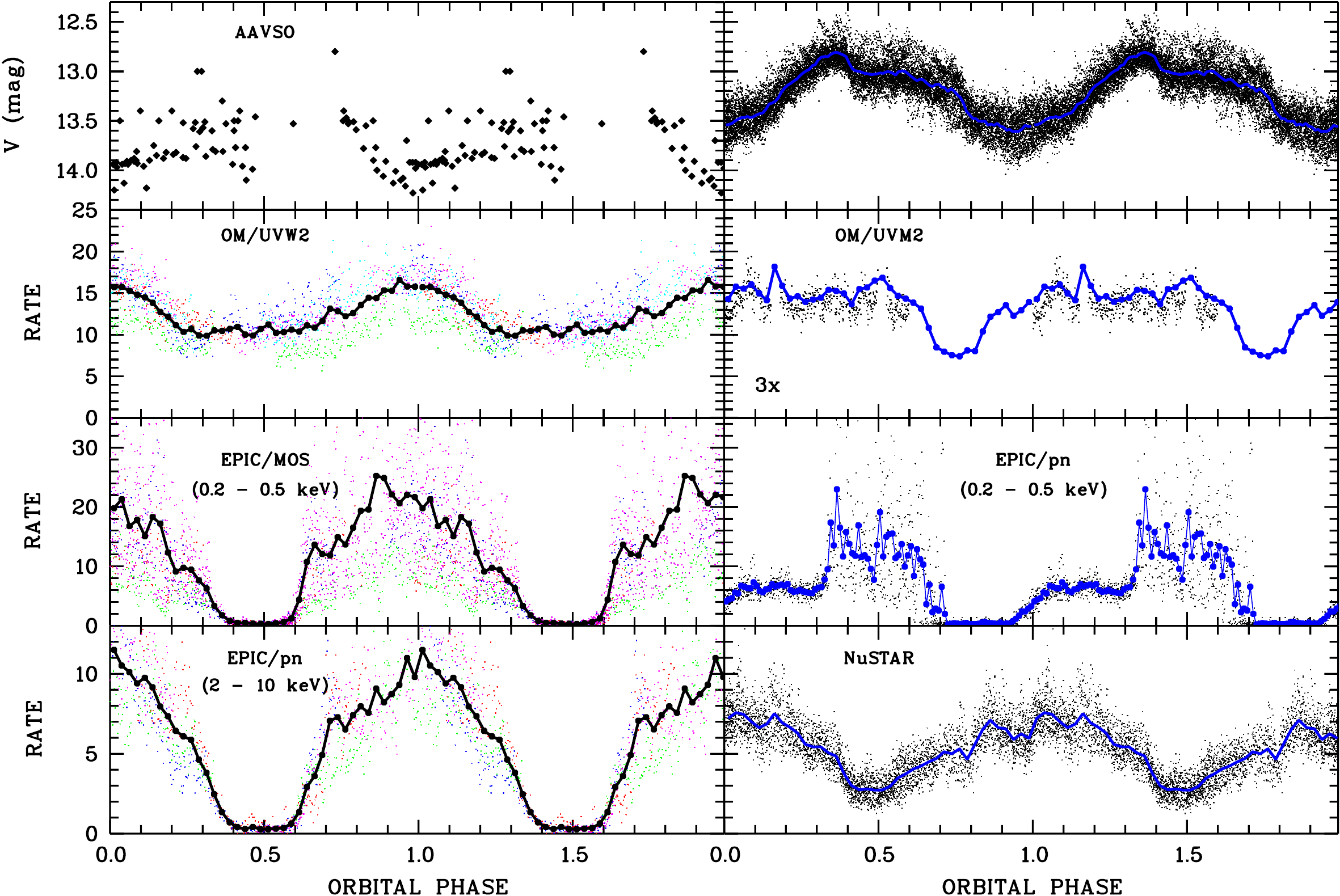}}
\caption{From top to bottom: Optical,
  ultraviolet, soft, and hard X-ray light curves obtained in 2005 (left) and
  2015 (right). They were obtained from the AAVSO data base, with the optical monitor
  and with EPIC onboard XMM-Newton, and with NuSTAR, respectively. 
  Original photon data were binned with a time bin size of
  30\,s. Phase-averaged light curves have 40 bins per orbital cycle. All data
  are shown twice for better visibility. The OM-filters were not the
  same in 2005 and 2015. The y-axis scale for 2015 have been compressed by a factor of 3. The soft X-ray light curves (third panel from above)
  were built using photons in the energy range $0.2-0.5$\,keV with EPIC-MOS
  (2005) and EPIC-pn (2015) with the thin and the thick filters, respectively (see
  Table~\ref{t:obs}). The hard X-ray light curves are based on EPIC-pn (2005)
  covering the energy range $2-10$\,keV and NuSTAR (2015, no energy
  selection). Colors of individual data points obtained in 2005 indicate different
  satellite orbits (2005 data, color - orbit): green - 1027, red - 1028, cyan
  - 1029, blue - 1030, magenta - 1031.
\label{f:lcs0515}}
\end{figure*}

\subsection{Time resolved optical observations}
\subsubsection{STELLA and AAVSO photometry}
A long-term AAVSO V-band light curve covering the period between January 1, 1999, and June 30, 2016, with launch and observation dates with \xmmn\ (blue) and Chandra (red) is displayed in Figure~\ref{f:aavso_99_16} \citep[all AAVSO data were kindly made available via their dedicated web-site aavso.org, ][]{kafka+20}. From there it becomes clear that the 2005
observations were luckily obtained in a brief active period whereas the 2015
observations fall in the center of an active period lasting almost 1.5 years.
Apart from sparsely sampled AAVSO data, the 2005 observations had no support
from ground-based telescopes.

The 2015 campaign was triggered and accompanied by comprehensive optical
photometry, mostly contributed by dedicated amateurs who shared their data via
the AAVSO web-page. Prior to March 2015 the amateurs reported about
one observation per night which indicated an active state. This was 
confirmed by time-resolved differential
photometric observations obtained with the STELLA-telescope 
equipped with WiFSIP \citep{strassmeier+04}. 
These were performed between March 27 and April 8 (see Tab.~\ref{t:stella} for details)
with a time resolution of 30\,s through a V-filter and revealed an orbital
variability between 13.8 mag and 12.8 mag. Based on the stability of the 
optical photometric light curves the satellite observations with NuSTAR and
\xmmn\ were triggered and eventually scheduled for April 6. The  
AAVSO amateurs were asked for supporting observations through BVRI filters in
the time period April 2 to April 16 (AAVSO alert notice 517). Further
STELLA-photometry was requested before, at, and after the date of the
satellite observations to monitor optical brightness changes with highest
possible cadence. 
The mean light curve of all STELLA V-band data is displayed in
Fig.~\ref{f:stella} and illustrates the large dispersion of the individual
data points around the mean behavior.

In response to the AAVSO alert notice, 35 AAVSO observers took part in the campaign. Between April 2 and 16 more than 15,700 individual measurements were reported. During the time interval when XMM-Newton and NuSTAR were observing, six AAVSO observers from the US, the UK, Spain and Canada could support the space-based observations with 1534 photometric data points. The phase-folded AAVSO data obtained in BVRI filters are displayed in Figure \ref{f:aavsofol}. By far the largest data set was obtained through the $V$-filter (more than 15,000 measurements), while of order 200 measurements were obtained each through $B,R,$ and $I$ filters. A primary orbital minimum was observed consistently between $\phi = 0.95 - 1.0$. The shorter the filter wavelength, the more symmetric the light curves become. The $I$ and $R$ light curves peak at phase $\phi = 0.3$, the $V$-band light curve at $\phi = 0.35$ and the $B$-band light curve at phase 0.4.

\begin{figure*}
\resizebox{\hsize}{!}{\includegraphics[clip=]{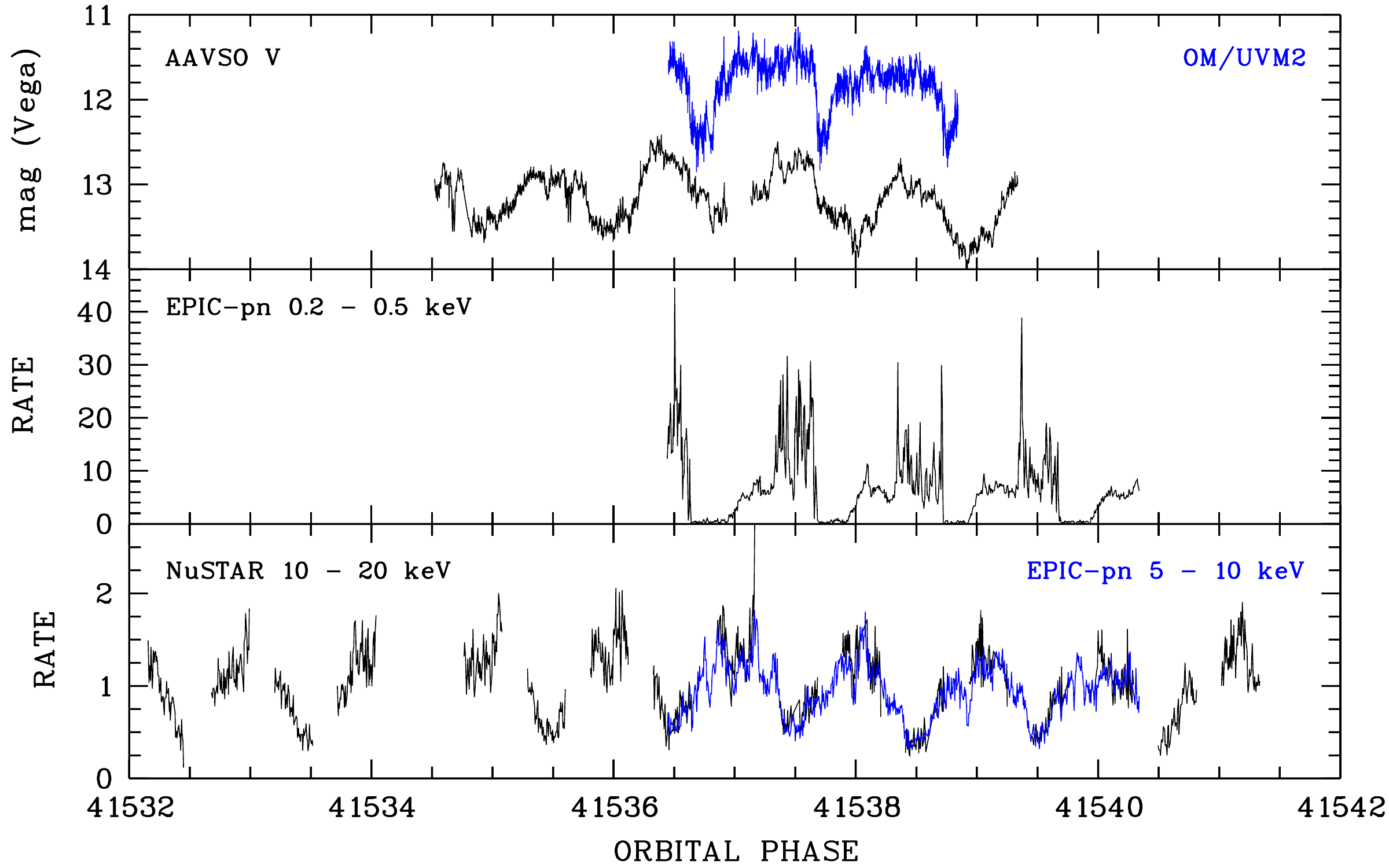}}
\caption{From top to bottom: Light curves obtained during the
  multi-observatory campaign in 2015 in original time sequence. Facilities
  used are indicated in the individual panels. Time bins are 10\,s for the OM, and 
  60\,s for EPIC-pn and NuSTAR, respectively.}
\label{f:lcs15}
\end{figure*}

\begin{figure*}
\resizebox{\hsize}{!}{\includegraphics[clip=]{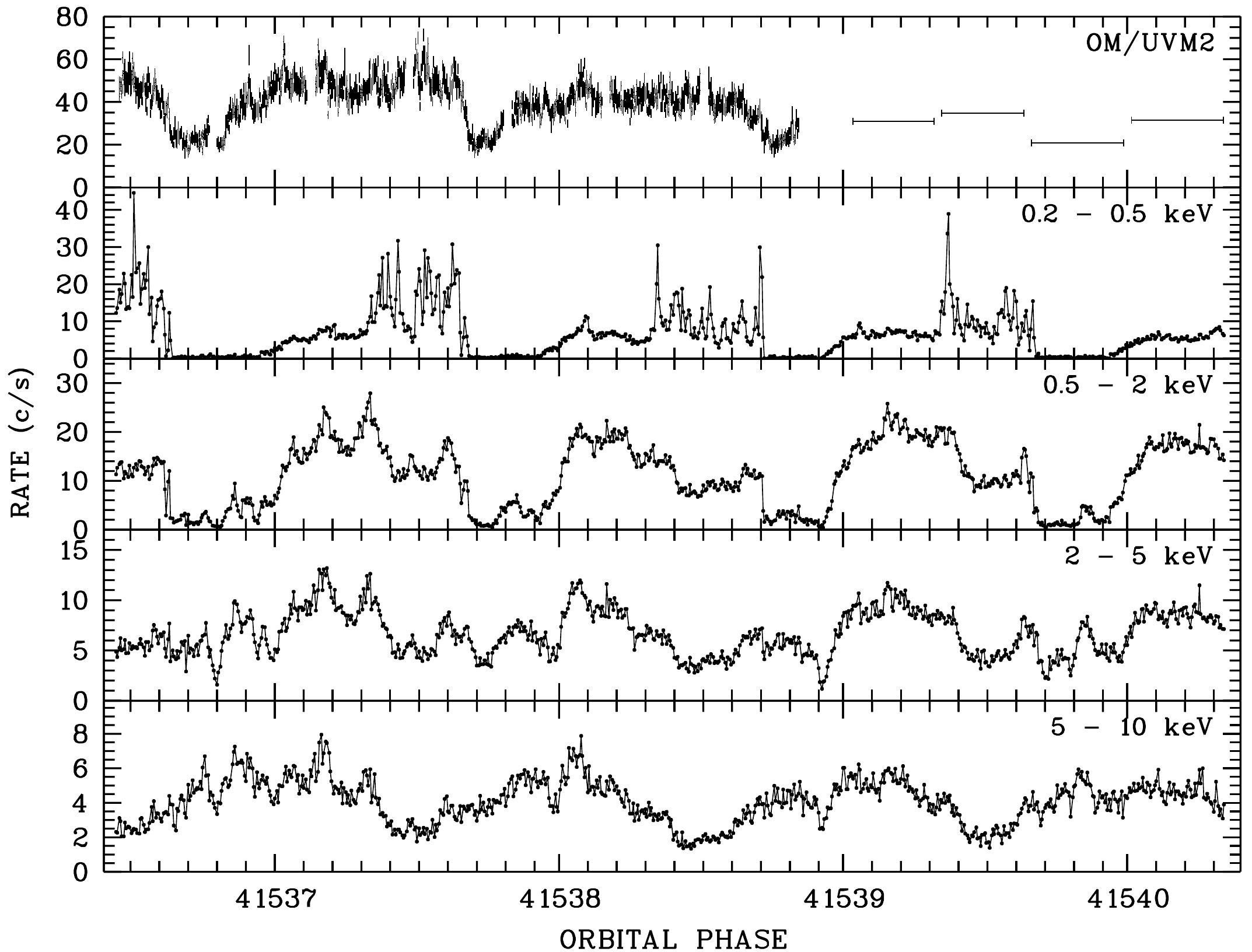}}
\caption{Energy-dependent EPIC-pn X-ray and OM UV light curves of the 2015 observation obtained with XMM-Newton. Time bins are 10\,s for the UV and 60\,s for the X-ray spectral range. The horizontal bars in the upper panel indicate the length and the mean brightness between 2200 and 2400\,\AA\, of the four grism spectra. Spectral flux density was converted to count rate using the conversion factor given in the XMM-Newton users handbook.}
\label{f:xlcs15}
\end{figure*}

\subsubsection{Spectroscopy with CAFE}
Following a Director Discretionary Time (DDT) request for ground-based spectroscopic coverage of our granted XMM-Newton/NuSTAR observations in 2015, observations were scheduled at the earliest possible convenience at the German-Spanish Astronomical Centre at Calar Alto (Spain). Observations were eventually conducted on the night of
April 18-19, 2015, with CAFE. CAFE is a fiber-fed Echelle spectrograph mounted at the 2.2m telescope covering the wavelength range 3960\,\AA -- 9500\,\AA\ with high resolution, $R\sim 62,000$.

Half a night was granted to our program and the observations were performed under a partially cloudy sky with nevertheless good transparency and median seeing of 1.5 arcsec well below the fiber entrance of 2.4 arcsec. The observations of AM Herculis began at UT 01:08:56 and lasted till UT 04:21:53 thus covering one complete orbital cycle. They were accompanied with calibration observations to perform bias- and flatfield corrections and wavelength calibrations. A total of 26 exposures were made on target with individual exposure times of 360\,s each.

Data reduction was done with ESO/MIDAS using the ECHELLE context. We performed standard reduction steps: subtraction of the bias level, cropping the images, order identification with the median master flat-field and wavelength calibration with the averaged master Th/Ar arc calibration image. The wavelength calibration has a final RMS of 0.04\,\AA.

\section{Analysis and results}
\subsection{Multi-wavelength light curves in the 2005 normal and the 2015 reversed mode\label{s:xlcs}} 
The photometric data obtained at optical, ultra-violet and
X-ray wavelengths during our campaigns in 2005 and 2015 are displayed
in Figs.~\ref{f:lcs0515}, \ref{f:lcs15}, and 
\ref{f:xlcs15}. Figure~\ref{f:lcs0515} shows phase-folded and averaged light curves
comparing the 2005 and 2015 observations while figures~\ref{f:lcs15} and 
\ref{f:xlcs15} are illustrating results of 
the more comprehensive 2015 campaign. These show the evolution of the light  
curves in original time sequence from the optical to the hard X-ray regime 
in Figs.~\ref{f:lcs15} and Fig.~\ref{f:xlcs15}. 

In 2005 \amh\ was encountered in its regular accretion mode with one
dominating accretion region (loosely referred to as the accreting pole) 
giving rise to a bright phase at UV, soft, and hard X-ray energies.
The satellite data display a large scatter around the average behavior. The
scattering amplitude is energy-dependent and stronger at soft than at hard
X-ray wavelengths. The mean X-ray light curves in hard and soft X-rays are similar in shape but are not identical.  The hard X-ray light curve is more symmetric with a maximum centered on phase zero (as the UV) whereas the soft X-ray light curve is more skewed. The rise to the bright phase is steeper than the fall, and the orbital maximum occurs at phase $\sim$0.85. The X-ray minimum phase lasts from phase 0.42 -- 0.58 but has no defined sharp edges. It is usually interpreted as self-eclipse of the accreting pole by the white dwarf itself. We note that the self-eclipse is not total, \amh\ is detected at a mean rate of 0.33\,\cps\ during that phase interval (EPIC-pn, $2-10$\,keV). The AAVSO data that were obtained occasionally seem to indicate a moderate high state, they remain about 0.5 mag below those obtained in 2015. 

In 2015 the source was encountered in its reversed mode of accretion. The 
light curves are fundamentally different compared to those in
2005. Only the hard X-ray light curve 
resembles that of the regular mode, although it appears skewed with a long
increase and more sudden decrease and with a much smaller pulsed fraction of 
$\sim$65\%.
In hard X-rays, the orbital minimum ("self-eclipse") starts at the same phase 0.42 as in 2005 but ends already at phase 0.52, hence the self eclipse is centered
on an earlier phase whereas the center of the bright phase (phases of half light on
ingress and egress) occurs at a later phase, $\phi = 0.03$. 
The NuSTAR ($10-20$\,keV) and the EPIC-pn ($5-10$\,keV) light curves are
almost carbon copies of each other (cf.~Fig.~\ref{f:lcs15}), and they certainly
originate from the same region and same emission process.

Strikingly different is the soft X-ray light curve in the softest channel
between 0.2 and 0.5 keV. It displays an interval of
intense soft flaring centered on phase $\sim$0.52 followed by an interval of
almost complete extinction of the softest X-rays. The onset of flaring seems
to be a stable feature at phase $\phi = 0.337\pm0.004$, whereas flaring
abruptly ends due to strong absorption at phases 0.642, 0.684, 0.721,
0.680 in cycles 41536 -- 41539, respectively. 
Flaring is seen only in the softest channel, below 0.5 keV, while 
the light curves at higher energies are nonetheless strongly affected by
absorption (see Fig.~\ref{f:xlcs15}). In the reversed mode, a second pole is
thought to accrete in a blobby accretion scenario \cite[see][]{heise+85,hameury_king88}. 
We find it difficult to uniquely determine the azimuth of this second
accretion region, because the end of the flaring phase could not be measured
due to soft X-ray absorption. The longest flaring phase was observed in cycle
41538 lasting $0.39\pm0.01$ phase units and being centered on phase 0.53. The
true center could however occur at a slightly later phase. 

In 2005 the UV light curve showed a bright-phase hump due to reprocessing of
high-energy radiation, reminiscent of former IUE and HST results
\citep{gaensicke+95,gaensicke+98,gaensicke+06}. There is nothing like that in 2015; the UV light curve appears approximately flat but is modulated
as the soft X-rays due to absorption. Three absorption dips were observed, but
the third  only partially. The FWHM of the two dips which were fully
covered was $\phi = 0.213$ and 0.12, respectively. It is worth noting that
the ingress into the dips at UV and soft X-ray wavelengths are not carbon
copies of each other due to strong flaring at soft X-rays which is not
present in the UV. 


\subsection{X-ray spectral analysis of AM Herculis in its 2005 regular mode of
  accretion}
We begin with an analysis of the data obtained in 2005 in the regular mode of accretion. This serves as a reference for the data obtained 2015 in the reversed mode. In the following subsections we characterize the spectra of the bright and faint phases, respectively, and the UV variability. The plasma emission lines detected with EPIC and the RGS in 2005 are analyzed jointly with the data obtained in 2015 in later sections. The spectral analysis presented here was performed with XSPEC \citep{xspec96} mostly with version 12.9.1n. 

\subsubsection{The X-ray bright phase}
Only photons in the phase interval $\phi = 0.72 -1.20$ were regarded for an
analysis of the bright phase for both EPIC and RGS.
Due to calibration uncertainties of the EPIC-pn spectra in timing mode, only EPIC-MOS and RGS data were analyzed jointly in the bright phase. The EPIC-MOS data, however, suffered from photon pile-up. To mitigate the pile-up effect on the spectra the central region of the target was excised from the analysis. The exclusion radius was determined experimentally by increasing the exclusion radius until the distribution of singles, doubles, triples and quadruples followed the theoretical expectations. 

A joint spectral model was developed for EPIC-MOS and RGS data that were spectrally binned to contain 25 photons per bin at least, so that $\chi^2$ optimization was applied. It turned out that the MOS and RGS data were discrepant below 0.5\,keV which was attributed to remaining effects of pile-up and perhaps optical loading. MOS data were thus excluded from the fits below 0.5\,keV so that the continuum at the lowest energies was determined by the RGS.

As usual for polars and AM Herculis in particular, a soft blackbody-like emission component is superposed on thermal plasma emission, modified by some cold neutral and some inter- or circumbinary absorption. The Fe-line complex was found to be resolved clearly into an  apparent triplet, the emission lines of hydrogenic and helium-like Fe at 6.9\,keV and 6.7\,keV, respectively, from the accretion plasma, and the fluorescence line from a scattering source at 6.4\,keV. 

Plasma emission plus reflection was modeled as the sum of a hot and a cold collisionally ionized plasma component (APEC in XSPEC terms). The fluorescence line was modeled with either a Gaussian or a Compton reflection component \citep[pexmon,][]{nandra+07} with similar results. 

In XSPEC notation the models used here were {\tt TBabs*pcfabs*(bbodyrad + constant*(apec + apec + gaussian))} (model A) and {\tt TBabs*pcfabs*(bbodyrad + constant*(apec + apec + pexmon))} (model B) with reduced $\chi^2_\nu = 1.085$ and $1.064$ for 2035 and 2034 degrees of freedom, respectively. Following \citet{beuermann+12} the column density of cold interstellar absorption (model component \texttt{TBABS}) was fixed at $5\times 10^{19}$\,cm$^{-2}$. The partial covering absorption component was found to be important to explain the hard thermal plasma spectra and thus avoid unrealistic high plasma temperatures. Without its inclusion, the hot plasma component hits its upper limit at $kT = 80$\,keV while still not giving a convincing representation of the data ($\chi^2_\nu = 1.26$). Abundances as given by \citet{angr89} were used for fitting the spectra. 


\begin{figure}
\resizebox{\hsize}{!}{\includegraphics[clip=]{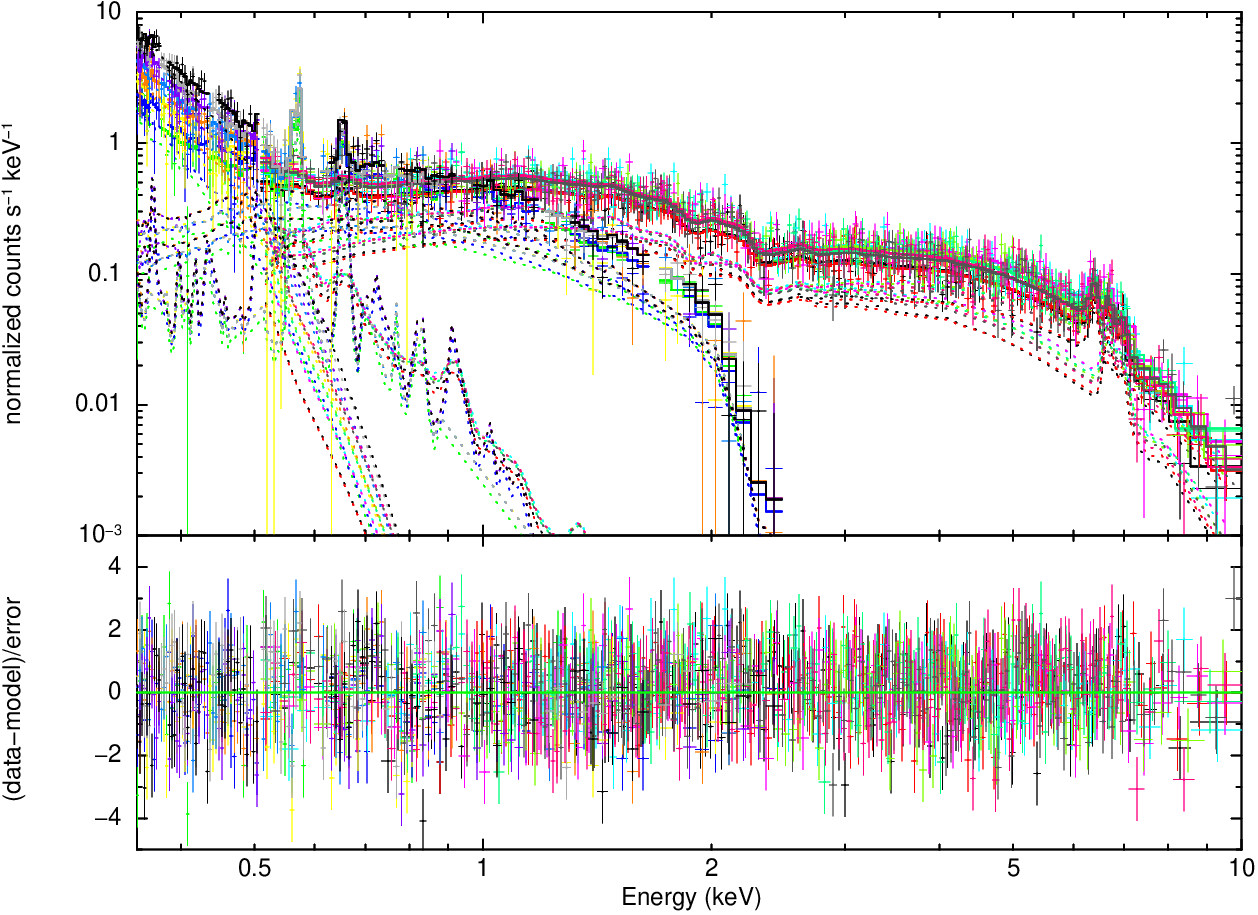}}
    \caption{Bright-phase EPIC-MOS and RGS spectra of \amh\ obtained in 2005 with
      best-fitting model B. The individual model components are shown with
      dotted lines. The lower panel shows the residuals with respect
      to the best fit in units of $\sigma$. Different colors were used to indicate data from the different instruments involved in the fit.}
    \label{f:bripha2005} 
\end{figure}

The following best-fit parameters were determined jointly for the four data groups representing the four individual observations with \xmmn: 
$kT_{\rm bb} = 35.2 \pm 0.7$\,eV, $N_H (\rm pcfabs) = 11.2\times 10^{22}$\,cm$^{-2}$, $kT_{\rm cool} = 0.161/0.174$\,keV (for models A/B), $kT_{\rm hot} = 14.0/8.5$\,keV (for a representation of the data and the fit see Fig.~\ref{f:bripha2005}). Other parameters like bolometric fluxes, the covering fraction and the soft X-ray emission area are given in Table \ref{t:fitbri05} for each individual \xmmn\ observation. The normalization parameter of the blackbody $N_{\rm BB}$ (in units of $R^2_{\rm km}/D^2_{10\rm kpc}$) was converted to a spot radius of an assumed circular emitting area at the distance to AM Herculis of 87.53\,pc \citep{Gaia2018, Bailer-JonesEtAl2018}.

The flux ratio $F_{\rm bb,bol}/F_{\rm X,bol}$ was derived without any further correction due to reflection or visibility. It indicates a mild to a pronounced soft X-ray excess which was found to be strongly correlated with the total X-ray flux, and, while the thermal flux was found to be roughly constant, the soft excess is correlated with the soft X-ray flux.

\begin{table*}
\caption{Bright and faint phase spectral parameters in 2005 for the model {\tt TBabs*pcfabs*(bbodyrad + constant*(apec + apec + gaussian))}. Flux units are given in $10^{-10}$ erg cm$^{-2}$ s$^{-1}$ and
$10^{-12}$\,\fint, for the bright and faint phases, respectively. The plasma flux $F_{\rm pl}$ is the sum of the bolometric fluxes of the cool and the hot components. CovFrac is the covering fraction of the partial covering absorber. The luminosity $L_{\rm X}$ is the sum of the blackbody and the plasma components with $L = \eta \pi d^2 F_{\rm bol}$ with geometry factors $\eta_{\rm BB} = 2$ and $\eta_{\rm pl} = 3.3$, respectively \citep{beuermann+12}. The mass accretion rate was computed for a white dwarf with $M_{\rm wd} = 0.78$\,\msun.   
\label{t:fitbri05}}
\begin{tabular}{r|crccrcc}
Rev. & $F_{\rm bb} $ & $R_{\rm BB} $ & CovFrac & $F_{\rm pl}$ & $F_{\rm bb} / F_{\rm pl}$ & $L_x$ & $\dot{M}$\\
& $[$cgs$]$ & $[$km$]$ & $[$\%$]$  & $[$cgs$]$  & & $[$erg s$^{-1}$$]$ & $[$ \msun/yr $]$\\
\hline
\multicolumn{6}{l}{\it (bright phase)} \\
1027 & $ 6.7\pm0.1$&  65 & 50 & $2.64\pm0.06$ & 3.4 & $5.1\times 10^{32}$ & $5.9\times 10^{-11}$\\
1028 & $19\pm2$    &  94 & 50 & $4.06\pm0.14$ & 4.7 & $12 \times 10^{32} $& $1.4 \times 10^{-10}$\\
1030 & $26\pm2$    & 113 & 61 & $3.68\pm0.09$ & 7.0 & $15 \times 10^{32}$ & $1.7 \times 10^{-10}$ \\
1031 & $33\pm2$    & 125 & 56 & $3.24\pm0.07$ &10.1 & $19 \times 10^{32}$ & $2.0 \times 10^{-10}$\\
\hline
\multicolumn{6}{l}{\it (mean faint phase)} \\
--- & $8.3^{+2.1}_{-1.1}$ & 6 & --- & $9.2^{+0.7}_{-0.6}$ & 0.9 \\
\hline
\end{tabular}
\end{table*}


The plasma temperature was also determined from just fitting a restricted 
energy range, 5.8 -- 7.5 keV, centered on the Fe-line complex. Thus we used the line strength of hydrogenic and helium-like iron as a bolometer. The model used writes {\tt APEC+BREMS+GAUSS} and revealed a plasma temperature of 10.3 keV, which we refer to as $kT_{\rm line}$.

The white dwarf in \amh\ has a mass of $M_{\rm WD} = 0.78^{+0.12}_{-0.17}$\,\msun\ \citep{gaensicke+06} implying a shock temperature of $kT_{\rm shock} = 35\pm12$\,keV. For an assumed specific mass flow rate $\dot{m} = 1$\,\mrats\ and a magnetic field of $B = 14$\,MG the maximum electron temperature $T_{\rm max,e}$ is about 40\% of the shock temperature, $kT_{\rm max, e} = 14$\,keV \citep{fischer_beuermann01} which is in good agreement with the hot plasma emission component of model (B). The temperature in the  line formation region $T_{\rm line}$ can be regarded a lower limit of $T_{\rm max,e}$ implying $kT_{\rm shock} = 26$\,keV. This value is lower than but still compatible with the expected value of 35 keV. 

\begin{figure}
\resizebox{\hsize}{!}{\includegraphics[clip=]{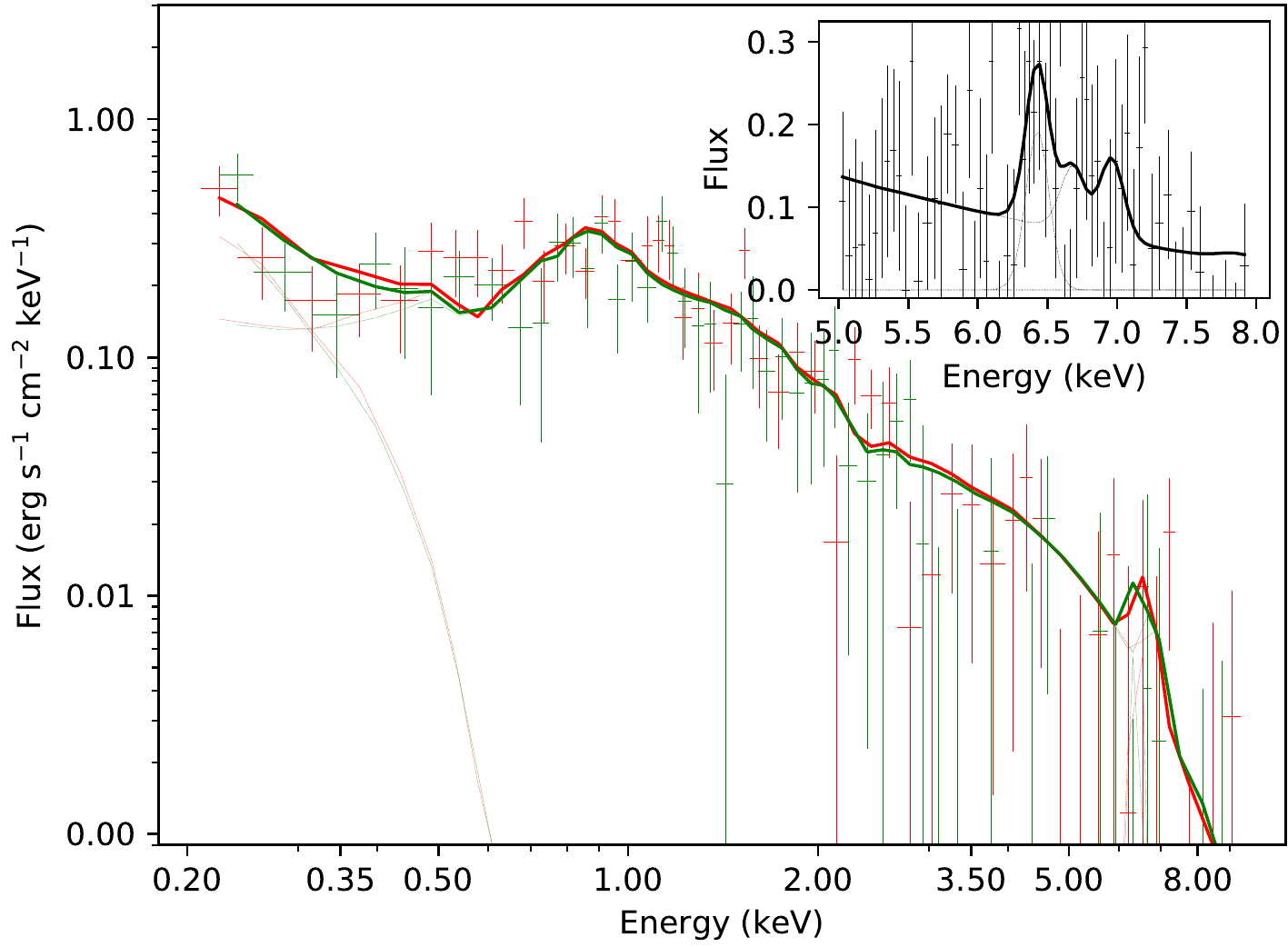}}
    \caption{Faint-phase EPIC-MOS (main panel) and -pn (inset) spectra of
        \amh\ obtained 2005 with best-fit model spectra overplotted. Observed
        data were binned with a minimum of 25 counts per spectral bin. Two
        model components are shown, the soft blackbody and the sum of all
        others originating from the accretion plasma
    \label{f:faint2005} }
\end{figure}

\subsubsection{The X-ray faint phase -- self eclipse}
\amh\ was sufficiently bright in the faint phase to perform a spectral analysis. Data from EPIC-MOS over the full spectral range and EPIC-pn just for the Fe-line complex in the restricted energy range $5.0-8.0$\,keV were used for this exercise. Spectra were generated using photons registered during the phase interval $0.42-0.57$ for each revolution of the spacecraft individually and combined into one
spectrum per instrument. 

The spectra were modeled with the same model as applied to the bright-phase data with the exception of the partial covering absorber to yield $kT_{\rm BB} = 36\pm10$ eV, 
$kT_{\rm cool} = 0.65\pm0.07 $\,keV, and $kT_{\rm hot} = 12^{+10}_{-5}$\,keV, respectively, where errors are given for a 99\% confidence interval. Derived parameters are also given in Tab.~\ref{t:fitbri05}, while the data and the best fit are shown in Fig.~\ref{f:faint2005}. Fitting was done with unbinned spectra (with a minimum of one count per bin), the data were later binned just for presentation in the mentioned figure.

The most important point to highlight here is the necessity to incorporate a soft component, parameterized as a blackbody, which hints to a second emission region on the white dwarf (second pole) as its source. It thus appears reasonable to assume that the thermal plasma component also originates form this second pole and is not scattered radiation of the primary accretion region. At the given distance to \amh\ the normalization of the blackbody reveals an emitting area of just a few
kilometers. 
It is interesting to note that the inferred blackbody temperatures in the bright and the faint phases are the same within the errors. Only the size of the secondary region is smaller than the primary region by large factors. No soft X-ray excess was found for the secondary region.

We also performed a fit to the EPIC-pn spectrum in the energy range $5 - 8$\,keV with a hot (12 keV) plasma emission model and (without) a Gaussian emission line  representing the 6.4\,keV line of neutral iron \citep[c.f.~][]{ishida+97}. The fit with the Gaussian line was clearly improved with a reduced
$\chi^2_\nu = 1.021$ (46 d.o.f.) compared to 1.188 (for 48 d.o.f.) for the model without the line but the model without the line cannot be rejected on safe grounds. We regard this result as weak evidence for the presence of the line. Its equivalent width is $0.7 \pm 0.2$\,keV (1$\sigma)$ which is higher than compatible with reflection but points to an unseen, scattered emitter as in, e.g., V902 Mon \citep{worpel+18}

\begin{figure}
\resizebox{\hsize}{!}{\includegraphics[clip=]{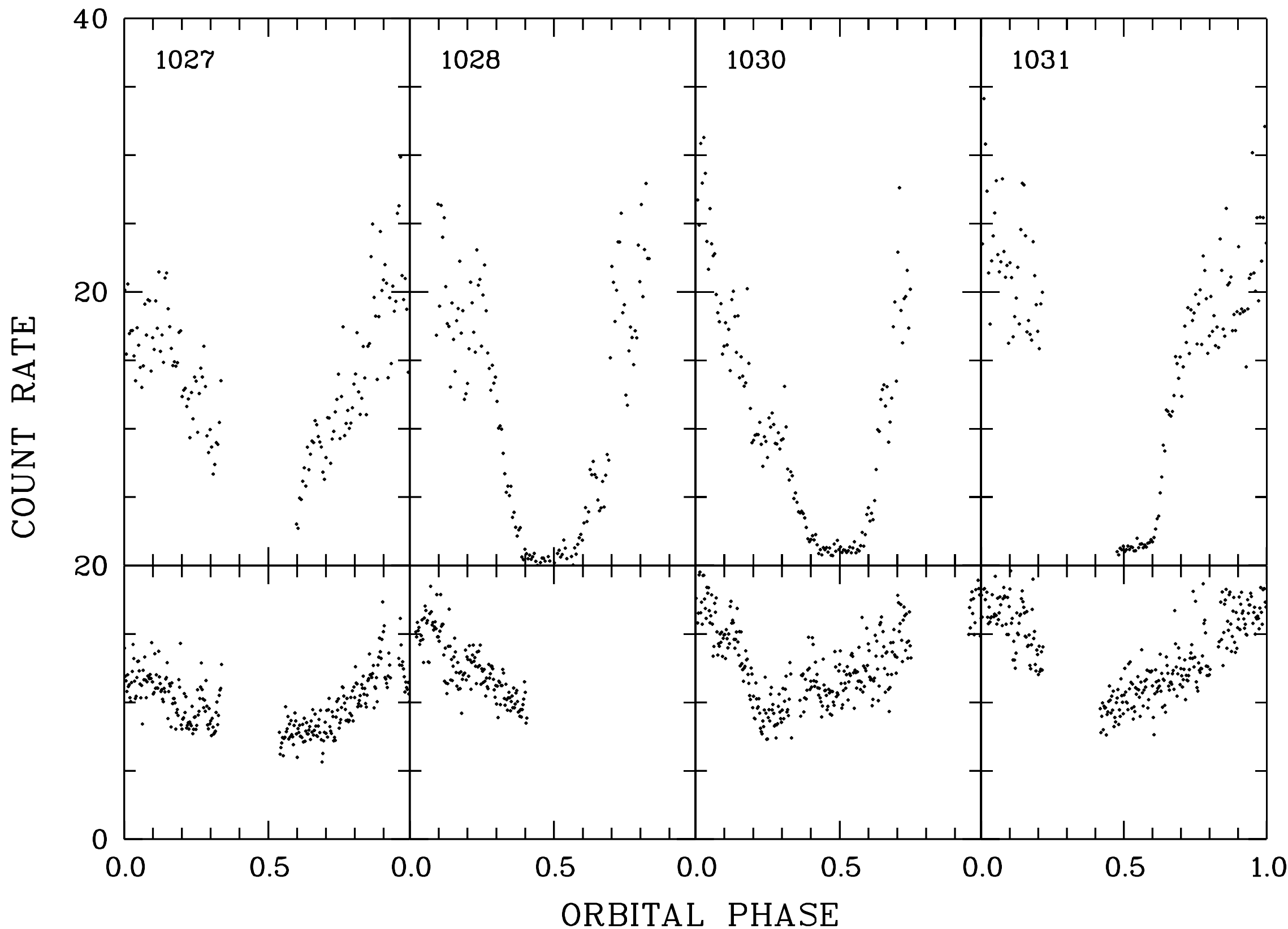}}
    \caption{Phase-folded Optical Monitor (OM, filter UVW2, lower panels) and EPIC-pn ($0.4-10$\,keV, upper panels) light curves per revolution obtained in 2005. The \xmmn\ revolution is indicated in the upper panels. 
    \label{f:om_pn_2005}}
\end{figure}

\begin{figure}
\resizebox{\hsize}{!}{\includegraphics[clip=]{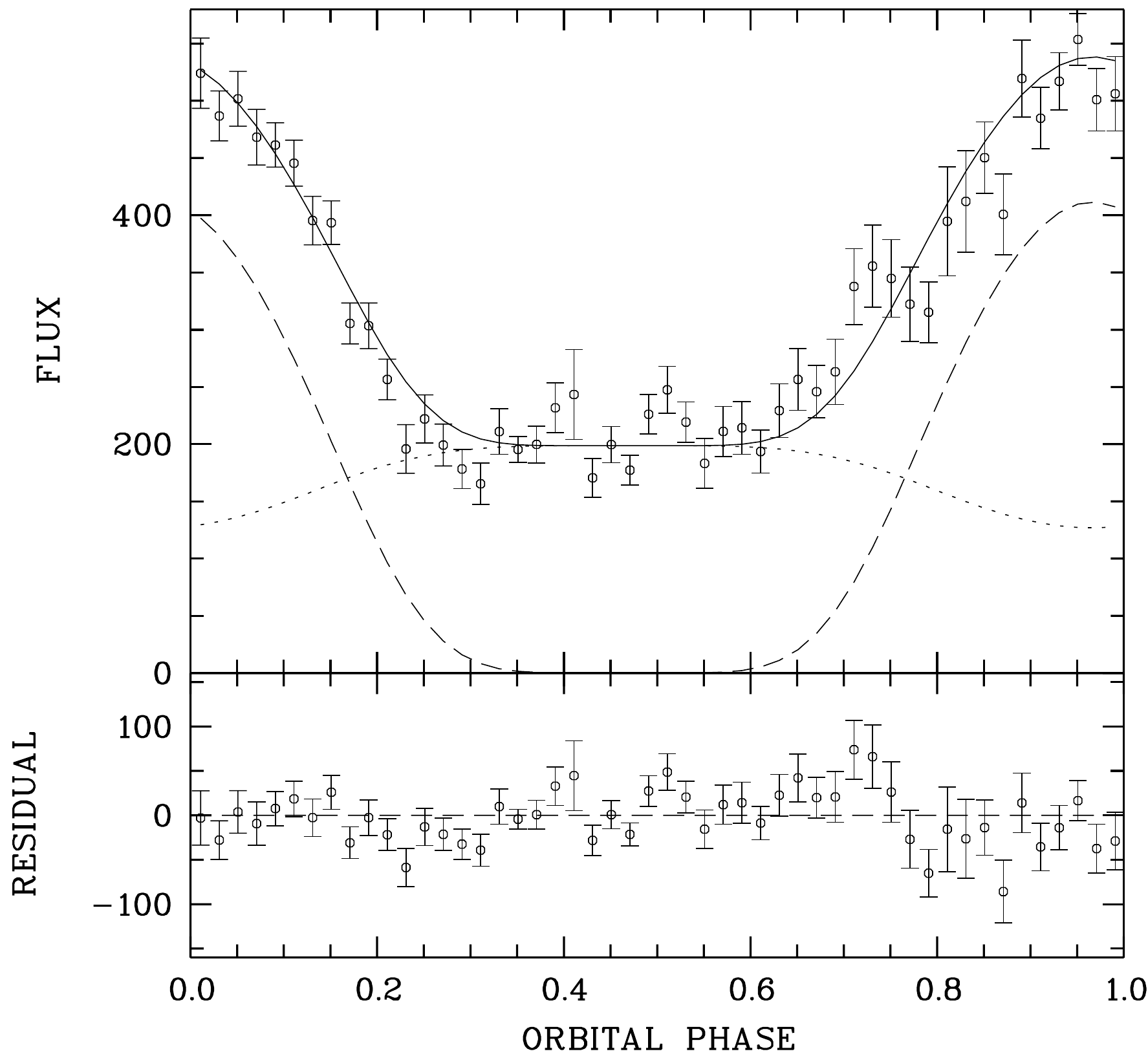}}
    \caption{Phase-averaged optical monitor light curve and spot model. Fluxes
      $F_\lambda$ are given in units of $10^{-16}$\,\flux. The best-fit is shown with a solid line, the contributions from the heated spot and the unheated white dwarf with dashed and dotted lines, respectively. 
    \label{f:spotmodel2005}}
\end{figure}

\begin{figure}
\resizebox{\hsize}{!}{\includegraphics[clip=]{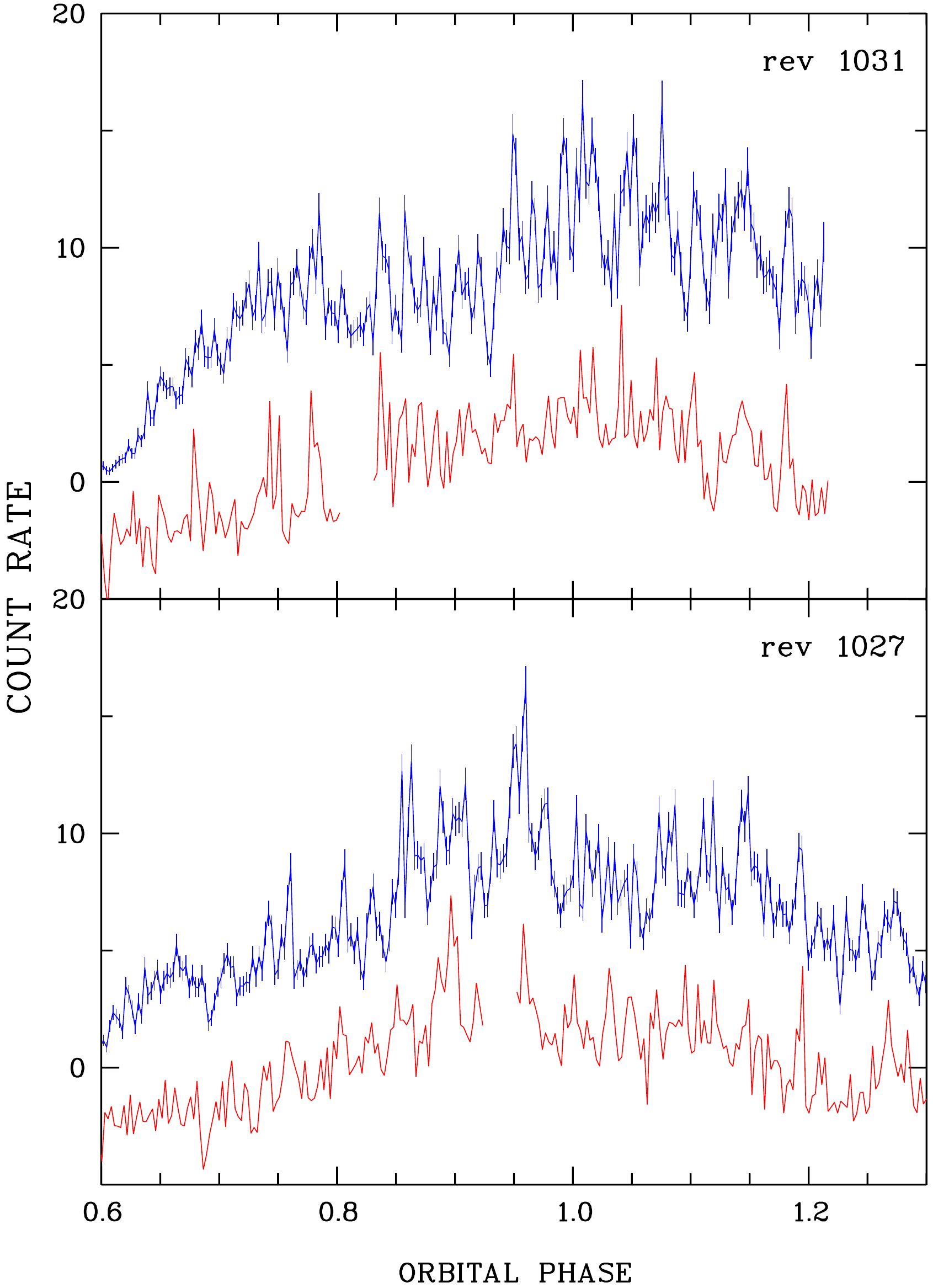}}
    \caption{Trains of binned photons observed in 2005 in the UV (OM/UVW2, red color)
      and the hard X-ray regime ($2-10$\,keV, blue color). Bin size is 30\,s
      each. The \xmmn\ revolution is indicated. The UV-data were plotted with
      an offset of $-5$ and $-8$ units in the lower and upper panels, respectively.
    \label{f:om_pn_2005_det}}
\end{figure}

\subsubsection{Variability in the UV\label{s:uv2005}}

In Fig.~\ref{f:om_pn_2005} the phase-dependent light curves in the UV
(2120\,\AA), and the hard X-ray regime are shown (0.4-10 keV), highlighting
similarities and dissimilarities of the light curves in the two wavelength
ranges per revolution. One observes a steady brightness increase in both
regimes at orbital maximum. The individual OM light curves do not display the
pronounced orbital minimum that is so remarkable at X-ray wavelengths. The
average UV light curve however displays the same shape as seen before with IUE,
HST, and FUSE \citep{gaensicke+95,gaensicke+98,gaensicke+06}. 

As shown in the aforementioned studies, 
the phase-dependent photometric variability 
can be explained by the projection of an accretion-heated spot. 
A similar model for a heated spot was applied here to describe the OM data. 
A grid of pure hydrogen atmosphere models was folded through the
response curves of the UV filter. A circular spot was placed on the
white dwarf and was characterized by its position, its extent and its
temperature distribution. A linear temperature decrease was assumed from a
maximum in the spot center to the temperature of the undisturbed white
dwarf. Since the data appeared rather noisy, several important parameters were
predefined. The distance was assumed to be $87.53\pm0.14$\,pc. The white dwarf has a mass of 0.78\,\msun\ and a
temperature of $T_{\rm eff} = 20000$\,K. Its radius was estimated using
\cite{Nauenberg1972}'s relation \citep{gaensicke+06}. 
The inclination was set to $i = 50\degr$. 

At phase 0.5 the accretion spot is not in view, 
hence the observed data at that phase are expected to contain only 
contributions from the undisturbed photosphere and the stream. 
The model gives a forecast for emission from the white dwarf at that phase which falls short of the observed brightness. 
The stream component was treated as a constant background at a level of  
$F_{\rm back} = 3.9 \times 10^{-14}$\,\flux\ \citep{gaensicke+95} and subtracted from the data shown in Fig.~\ref{f:spotmodel2005}.

An optimum fit was determined via $\chi^2$ minimization. The data, the model
and the residuals are shown in Fig.~\ref{f:spotmodel2005}. We find a
colatitude of 65\degr\ \citep[of course very much dependent on the assumed
inclination, see the discussion in ][]{gaensicke+01}, a spot azimuth of about
12\degr, a spot opening angle of 35\degr\ 
and a maximum spot temperature of 66,000\,K. The extra flux originating from
the accretion-heated spot is $F_{\rm UV} = 0.9 \times 10^{-10}$\,\fint, the
spot covers a fractional area $f\simeq0.09$ of the white dwarf's surface.

Binned hard X-ray ($2-10$\,keV) 
and UV light curves obtained in revolutions 1027
and 1031 are shown in original time sequence in
Fig.~\ref{f:om_pn_2005_det}. The chosen X-ray band covers the thermal emission
component from the main accretion spot. Both light curves display strong
flickering behavior. Some, not all, of the hard X-ray flares show a 
corresponding signal in the UV. Hard X-ray flares are due to some kind
of non-stationary accretion. The observation of a 
correlated UV-signal is a direct proof of instantaneous re-processing
of radiation in the photosphere of the WD.
We underline that the reprocessing 
component is detected in the UV, not in soft X-rays, lending support to G\"ansicke's earlier claim based on energetic grounds only \citep{gaensicke+95}.
The lack of energy resolution in the UV prevents us from drawing more detailed conclusions as far as the energy budget is concerned.

\begin{figure}
\resizebox{1.03\hsize}{!}{
\includegraphics{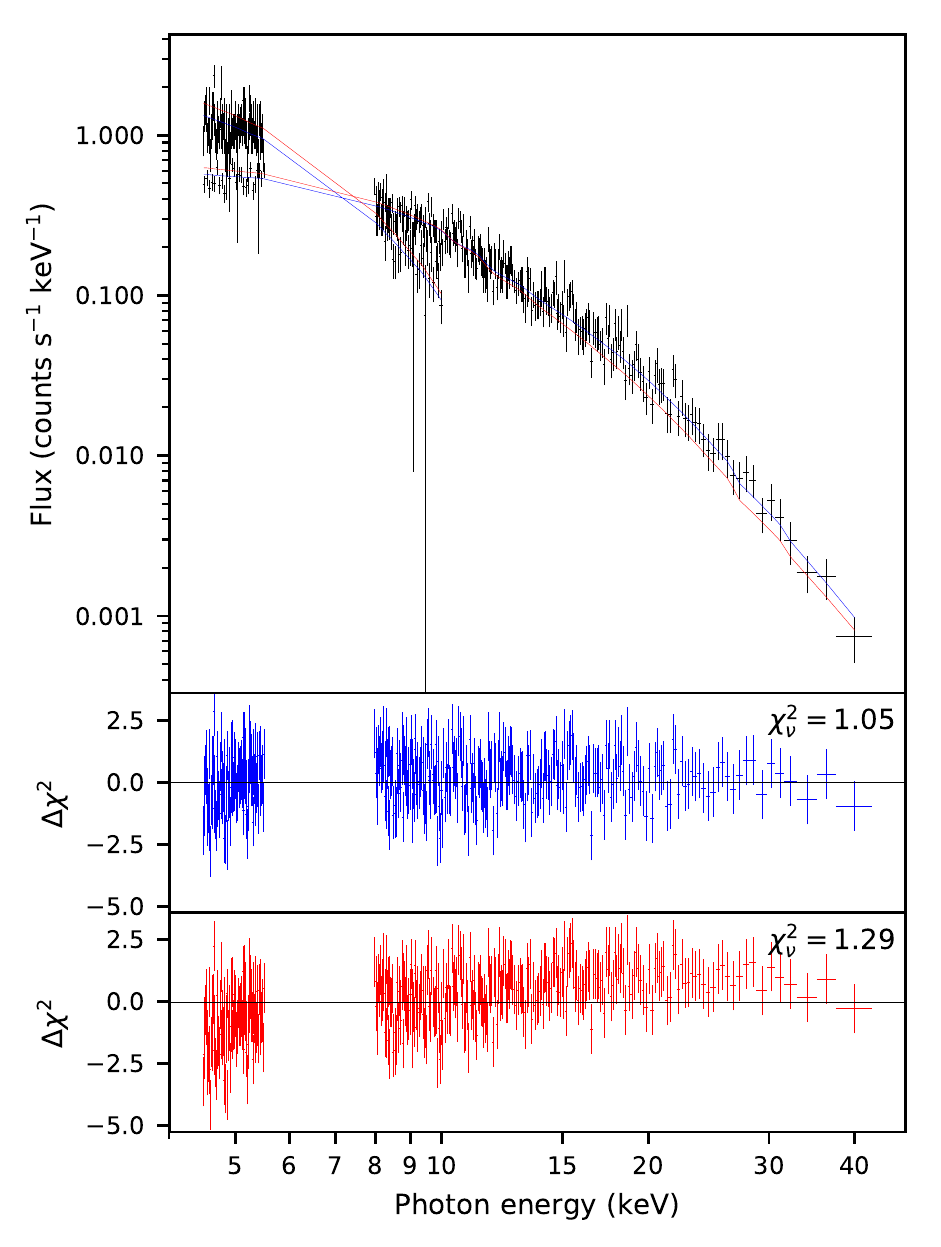}}
    \caption{Combined \xmmn and NuSTAR bright-phase spectrum obtained in 2015. Model fits and residuals with (blue) and without (red) reflection component are shown in the two lower panels.}
    \label{f:reflection} 
\end{figure}

\subsection{Spectral analysis of AM Herculis in its 2015 reversed mode of accretion}

In this subsection the main spectral characteristics of the 2015 X-ray data are analyzed. We begin with a joint fit to EPIC-pn, EPIC-MOS and NuSTAR data to determine the reflection continuum, both in the mean and in phase-resolved spectra. We then fit the full bright-phase spectrum including the lines and use this parametrization as background to model the soft blobby accretion phase. Finally we strive for an understanding of the X-ray and UV absorption dips. 

\begin{figure}[t]
    \includegraphics{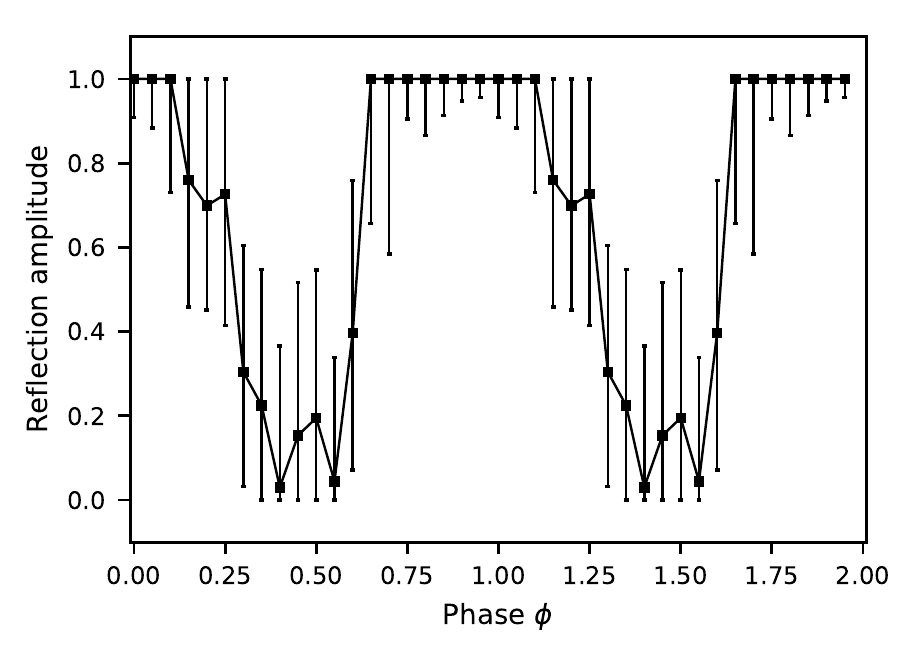}
    \caption{ Phase-dependent reflection amplitude for EPIC and NuSTAR
      spectra. Two cycles are plotted for clarity, and  
              error bars indicate the 1$\sigma$ uncertainty. }
    \label{f:phas_reflection} 
\end{figure}

\subsubsection{The combined XMM-Newton and NuSTAR reflection spectrum}

We extracted photons during the phase interval 0.05$<\phi<$0.25, called the plateau phase. For EPIC-pn only photons detected as singles and doubles were used and extracted in separate spectra. Following \cite{MukaiEtAl2015} we discarded data below 4.5 keV and between 5.0 and 8.0 keV, thus excluding the iron line complex and other emission lines. A bremsstrahlung emission model with fixed temperature at 15\,keV was used, modified with the continuum reflection model {\tt reflect}.
This we fitted twice, once with the reflection amplitude allowed to vary and once with the reflection amplitude fixed to 0.0 but bremsstrahlung normalization still variable. We observed that the reflection component gives an improvement of $\Delta\chi^2=437$ for 1411 degrees of freedom and a reflection amplitude of around 0.8, a very significant detection of reflection (Figure~\ref{f:reflection}).
Leaving the bremsstrahlung unconstrained gives a temperature of $14.5$\,keV without significantly affecting the reflection amplitude.

We then searched for the phase-dependent behavior of the reflection continuum. Therefore, we divided all spectra into twenty phase bins and fitted them jointly with a single temperature bremsstrahlung modified by reflection. The bremsstrahlung temperature was kept fixed at 15\,keV and its normalization allowed to vary. The reflection component parameters were kept fixed at their default values except for the amplitude, and each spectral group was given its own multiplicative calibration constant. The maximum physically plausible reflection amplitude is unity, representing a flat infinite plane, so we capped this parameter at 1. Thus, the XSPEC model is \texttt{const*reflect*bremss}. The $\chi^2_\nu$ for all twenty fits were around unity, indicating that there is no need to add more model components (Figure~\ref{f:phas_reflection}). We found clear evidence of phase-dependent reflection, although with large uncertainties with a maximum of one on the rising branch of the hard X-ray light curve and a minimum, consistent with zero, which coincides roughly with the soft flaring interval of the light curve. 

The reflection amplitude for  spectra in the phase interval $0.35<\phi<0.55$ appears consistent with zero. To verify, we extracted mean spectra for this phase interval and fitted it as above. With a greater S/N, we found a reflection amplitude of $0.074^{+0.062}_{-0.060}$ to $1\sigma$ significance, still compatible with zero.

\subsubsection{The 2015 prime pole X-ray bright phase\label{s:2015bri}}
\begin{figure} 
\resizebox{\hsize}{!}{\includegraphics{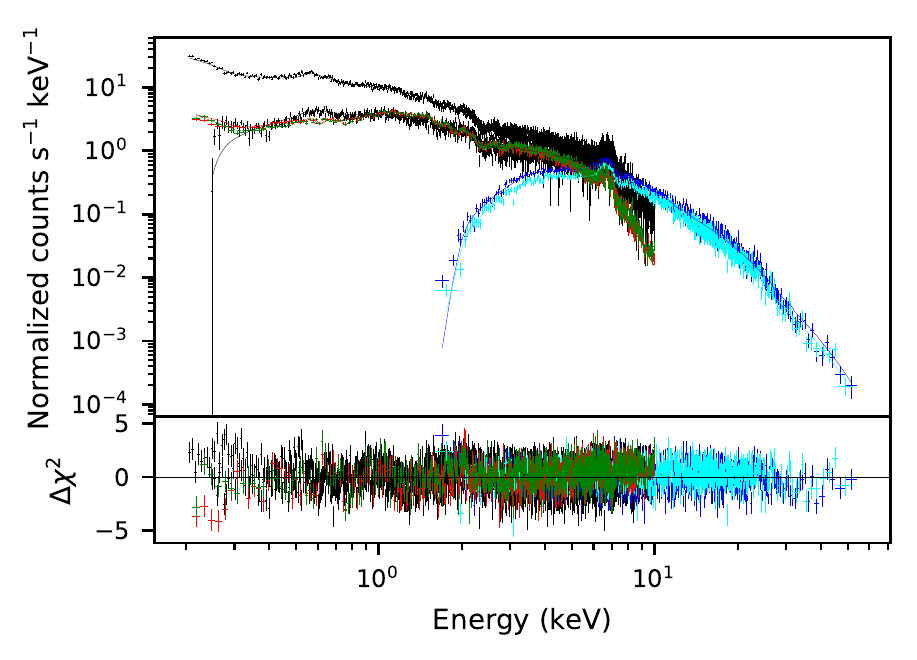}}
    \caption{Bright-phase spectrum (plateau)
    from 2015 with best fitting model.     }
    \label{f:plateau_fit} 
\end{figure}
Obtaining a satisfactory fit to the plateau spectrum (phase 0.05--0.25) over the entire energy range was difficult, and we proceeded in stages. We first constrained the plasma temperatures using the iron lines, fitting the energy range 6.0-7.2\,keV with a reflected two-temperature APEC model plus a Gaussian near 6.4\,keV. The EPIC-$pn$, EPIC-MOS, and NuSTAR instruments were fit jointly, with a multiplicative factor for each instrument to account for cross-instrument calibration. We also split the EPIC$-pn$ photons into single patterns and multiple patterns to correct for instrumental artifacts at low energies. Additionally, the redshifts of the reflection and APEC components were allowed to vary between instruments, to account for other calibration issues. The Gaussian's energy was also tied to the redshift.
With the redshifts and APEC temperatures found, we extended the energy range of the fits upward to 10\,keV for the EPIC instruments and 55.0\,keV for NuSTAR. There were very few photons above these energies. 

To fit this spectrum adequately we had to add a third APEC component, a soft blackbody, a partially covering local absorber, and the $5\times10^{19}$\,cm$^{-2}$ interstellar absorber. That is, our XSPEC model was {\tt const*tbabs*pcfabs*reflect*\\(bbodyrad+apec+apec+apec+gaussian)}. The resulting fit matches the continuum and iron line complex well, with a $\chi^2_\nu$ of 1.18 for 5145 degrees of freedom. There is an unidentified absorption-like feature near 1.3\,keV, some kind of emission feature at 2.0\,keV, and some discrepancies at the lowest energies that lead to a not fully satisfactory fit  (see Figure \ref{f:plateau_fit}). Experiments with various kind of standard absorber models to explain the feature at 1.3 keV remained inconclusive. This feature might find an explanation in yet to be formulated complex absorbing structures involving, e.g., a partial covering warm absorber. 
We think, nevertheless, that we have correctly characterized the important features of the spectrum. The fit parameters are summarized in Table \ref{t:plateau_fit}.

\begin{table}
\caption{Parameters and fluxes for the plateau phase ($0.05<\phi<0.25$, top part) and the additional soft flux in the flaring phase (bottom part). Fluxes are bolometric, in units of $10^{-10}$\,\fint and all given uncertainties are at the $1\sigma$ level.}
\begin{tabular}{lr}
\hline
Local absorber $n_H$     & $(5.94\pm 0.14) \times 10^{22}$\,cm$^{-2}$ \\\vspace{1 mm}
    ... covering fraction    & $42.8 \pm 0.6$\% \\
    Reflection amplitude     & $0.71_{-0.09}^{+0.07}$ \\
    Blackbody kT$_{\rm BB,plateau}$   & $30.0^{+1.0}_{-3.1}$\,eV \\
    APEC $kT_1$              & $161\pm 3$\,eV \\
    APEC $kT_2$              & $5.0^{+0.2}_{-0.4}$\,keV \\
    APEC $kT_3$              & $15.7^{+0.2}_{-0.3}$\,keV \\
   $f_{\rm BB}$  & $6.97\pm0.11$\\
   $f_{\rm thermal}$  & $4.17\pm0.01$\\
   $L_{\rm BB+thermal}$ & $6.35 \times 10^{32}$\,erg s${-1}$ \\
   $\dot{M}$ & $7.37 \times 10^{-11}$\,\msun yr$^{-1}$\\
    \hline
     Blackbody kT$_{\rm BB, flare}$   & $33.7 \pm {0.1}$\,eV \\
$f_{\rm BB, flare}$ & $14.85\pm 0.08$ \\
   $L_{\rm BB, flare}$ & $6.81 \times 10^{32}$\,erg s${-1}$ \\
   $\dot{M}$ & $7.90 \times 10^{-11}$\,\msun yr$^{-1}$\\
\end{tabular}
\label{t:plateau_fit}
\end{table}

\subsubsection{Soft blobby accretion at the second pole}
\label{s:flaring}
\begin{figure}[t]
    \resizebox{\hsize}{!}{\includegraphics[angle=-90,clip=]{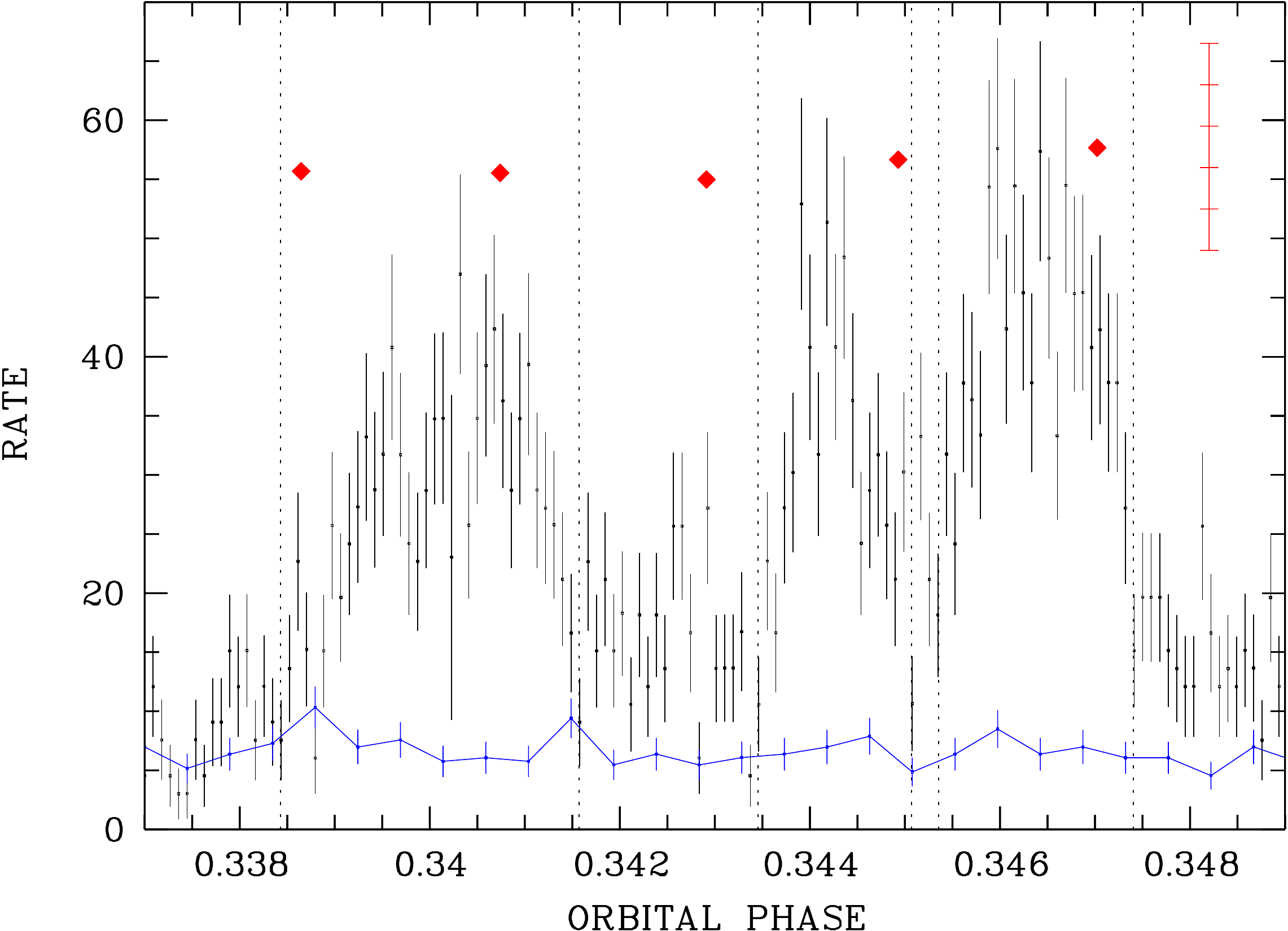}}
    \caption{Flare details in cycle 41538. The soft X-ray light curve (0.2-0.5\,keV) is shown in black with 1\,s time resolution, the hard X-ray light curve (2-5 \,keV) at 5\,s resolution in blue. Optical photometry is shown in red, the inserted scale bar ranges from $V=13.1$ at bottom to 12.6 mag with a 0.1 mag separation between ticks. Three flares are identified and marked by vertical dashed lines.
     \label{f:flarelc}}
\end{figure}

In this section we firstly describe attempts to fit the mean spectrum of the flaring interval followed by an analysis of individual flares. 
To fit the mean spectrum in the middle of the flaring phase, $\phi= 0.40 - 0.55$, we assume that the contribution from the main accreting pole, as characterized in the previous section, varies only in intensity, not in spectral shape, as the pole moves in and out of view. The flaring events are described as a soft blackbody superimposed on it, that is, a second blackbody in addition to the one already present in the plateau spectrum. 
We also assume that any non-flaring contribution from the secondary pole is negligible, an assumption we can justify from the observation that no brightness increase during a flare is seen neither at optical, nor UV or X-ray wavelengths (see Fig~\ref{f:flarelc} for an example at high resolution).
The temperature and normalization of the second blackbody were found to be slightly higher than that from the plateau phase, $kT=33.7\pm0.1$\,eV). 
The average bolometric flux of the additional blackbody, without reflection or absorption, was $1.485\pm 0.008\times 10^{-9}$\,\fint. The X-ray flux from the soft component of the second pole is higher than the summed X-ray flux (thermal plus blackbody) from the primary pole. Thus, the sum of the fluxes from both poles is $2.600\pm0.008\times 10^{-9}$\,\fint, which equates to a mean mass accretion rate of $1.53 \times 10^{-11}$\,\msun\,yr$^{-1}$. To derive this we used the same geometry factors as for 2005 (see Tab.~\ref{t:fitbri05}). The derived total mass accretion rate is within the range of rates covered by the values derived for just the main pole in the 2005 high state. 

To investigate the properties of individual flares and of the underlying
accretion filaments we inspected the EPIC$-pn$ light curve in the
energy range 0.1-0.5\,keV binned to 1 s by eye using TOPCAT \citep{Taylor2005}, and
identified 84 distinct events (see Figure~\ref{f:flarelc} for an example). We determined the start and end times of individual flares, the duration $\Delta t_{\rm flare}$, and the time of the flare maximum. The rise and fall times, $t_{\rm rise}$ and  $t_{\rm fall}$, were calculated as differences between the start and the end of a flare and the flare maximum. Start and end times were used to generate GTIs (Good Time Intervals) and \xmmn\ EPIC spectra were extracted for each flare individually. They were binned always with a minimum of 16 counts per bin. Then we fit each flare individually, from 0.2 to 10.0\,keV, using the model developed in the previous section as background plus a soft blackbody with variable temperature and normalization describing the flare proper. For the background spectrum all spectral parameters were held fixed but a variable multiplicative factor was applied to only allow the flare background to vary in intensity, and the covering fraction was left free.

\begin{figure}[t]
    \resizebox{\hsize}{!}{\includegraphics[clip=]{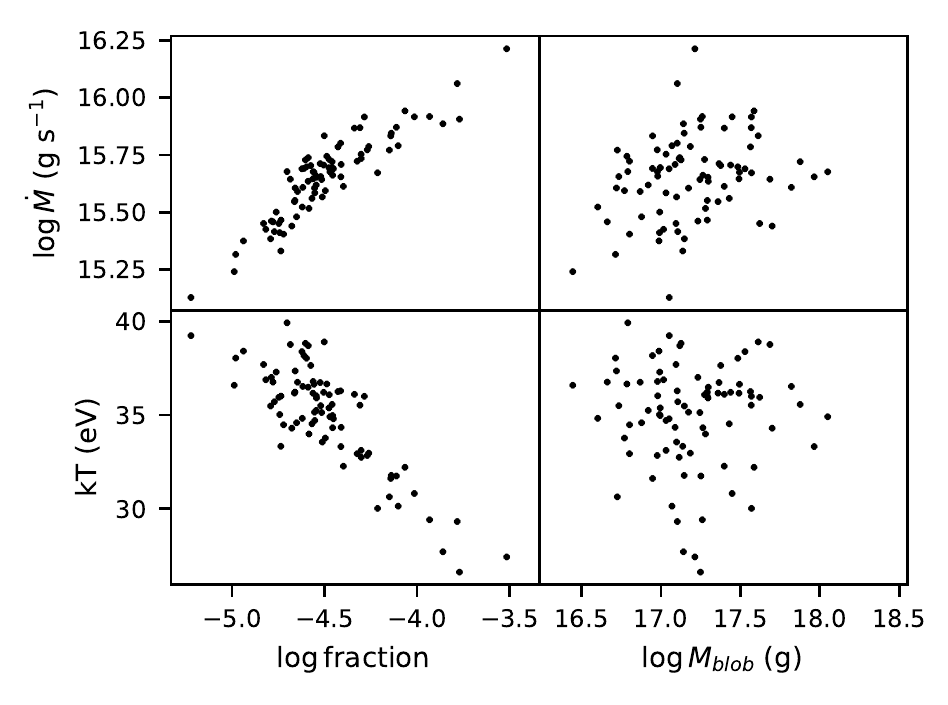}}
    \caption{Flares at the second pole of AM Herculis in its anomalous
      state. Shown are the mean mass accretion rate and temperature of the individual flares against the emitting fraction of the WD surface ({\it left panels}), and against the mass of the accreting blob ({\it right panels}). Although $\dot{M}$ and $kT$ are strongly dependent on the size of the emitting area, they appear to be independent of the mass of the infalling blob.
     \label{f:flarefits}}
\end{figure}

The results are summarized in Table~\ref{t:flare} (mean and extreme values and dispersion for all flare properties) and illustrated in Figure \ref{f:flarefits} for individual flares. Given the robustness (or simplicity) of the approach the fits were found to be satisfactory with a mean $\chi^2_{\rm red} = 1.28$, although not of sufficient quality in any individual case. Based on those fits a number of parameters were derived describing the radiation properties and energetics of the accreted filaments: the mean bolometric flux $F_{\rm BB}$, the fluence 
$\mathcal{F}_{\rm BB} = F_{\rm BB} \times \Delta t_{\rm flare}$, 
the bolometric luminosity $L_{\rm bol,BB}$, the emission area $A_{\rm BB}$ and the emission radius of an assumed circular radiating spot $R_{\rm BB}$, the radiating fraction $f_{\rm BB} = A_{\rm BB} /
(4\pi R^2_{\rm WD})$, the specific luminosity $L_{\rm f} = L_{\rm bol,BB} / f_{\rm BB}$, the mass accretion rate $\dot{M}$, the specific mass accretion rate (or mass flow rate) $\dot{m} = \dot{M} /  A_{\rm BB}$, an upper limit to the mass of the accreted filament $m_{\rm blob} = \dot{M} \times t_{\rm rise}$ and its length $l_{\rm blob} = \Delta t_{\rm flare} \times v_{\rm ff}$. The derived blob mass is regarded an upper limit since the value given depends on the observed photometric signal and not on any observable of the filament itself (the impact time is likely shorter than the photometric rise time). Similarly one may argue that the derived blob length is an upper limit because it is based on the observed duration of a flare. The derived average specific mass flow rate, $\langle\dot{m}\rangle = 22$\,g\,cm$^{-2}$\,s$^{-1}$, is certainly a lower limit to its real value. Its value was derived using the emission area of a filament while the accretion area is certainly much smaller. With $\dot{m} > 22$\,g\,cm$^{-2}$\,s$^{-1}$, we are very much in the regime of filamentary blobby accretion where hydrodynamic shocks are buried deeply below the photosphere of the ambient medium \citep{kuijpers_pringle82,frank+88,king00,litchfield_king90}.

\begin{table*}
\caption{Flare parameters derived from blackbody fits to 84 individual soft
  flares from the second pole in AM Herculis. Given are the mean values, the
  standard deviation, the minimum and the maximum values.\label{t:flare}}
\begin{tabular}{ccrrrr}
\hline
Parameter & Unit & Mean & StD & Minimum & Maximum \\ 
\hline
$\Delta t_{\rm flare}$ & s & 46.4 & 43.1 & 9 & 236 \\
$t_{\rm rise}$ & s & 7.5 & 10.5 & 1 & 86 \\
$t_{\rm fall}$ & s & 9.6 & 9.3 & 1 & 72 \\
$kT_{\rm BB}$ & eV &  35.0  &   2.8 &  26.6&  39.9 \\
$F_{\rm BB}$ & erg cm$^{-2}$\,s$^{-1}$ & $1.53 \times 10^{-9}$ &  $ 6.9 \times 10^{-10}$ &  $ 4.2 \times 10^{-10}$ &   $5.1 \times 10^{-9}$\\
$\mathcal{F}_{\rm BB}$ & erg cm$^{-2}$ &$ 6.39 \times 10^{-8}$ &  $ 5.84 \times 10^{-8}$ &  $ 8.7 \times 10^{-9}$  &   $ 3.5 \times 10^{-7}$\\
$L_{\rm bol,BB}$ & erg\,s$^{-1}$ &$ 7.0 \times 10^{32}$  &  $ 3.2 \times 10^{32}$  &  $1.9 \times 10^{32}$  &   $2.3 \times 10^{33}$\\
$A_{\rm BB}$ & cm$^{-2}$ &$ 2.8 \times 10^{14}$ & $ 2.8 \times 10^{14}$&  $ 3.9 \times 10^{13}$  & $2.0 \times 10^{15}$  \\
$R_{\rm BB}$ & km & 87     &   34   &   35        &       252\\
$f_{\rm BB}$ & -- &$ 4.2 \times 10^{-5}$ &  $ 4.2 \times 10^{-5}$ &   $ 6.0 \times 10^{-6}$ &  $ 3.1 \times 10^{-4}$\\
$L_{\rm f}$ & erg\,s$^{-1}$ &$2.1 \times 10^{37}$ &  $ 5.9 \times 10^{36}$ &   $ 6.8 \times 10^{36}$ &  $ 3.4 \times 10^{37}$ \\
$\dot{M}$ & g\,s$^{-1}$ &$ 4.9 \times 10^{15}$ &  $2.2 \times 10^{15}$ &  $1.3 \times 10^{15}$&  $1.6 \times 10^{16}$\\
$\dot{m}$ & g\,cm$^{-2}$\,s$^{-1}$ &22   &    6   &     7&   36 \\
$m_{\rm blob}$ & g  &$ 2.1 \times 10^{17}$ & $ 1.9 \times 10^{17}$ &   $ 2.8 \times 10^{16}$&  $1.1 \times 10^{18}$\\
$l_{\rm blob}$ & km &40300  &     56000 & 5300&  460000 \\
$l_{\rm blob}$ & $R_{\rm WD}$ &5.6   &   7.8  &    0.74  &  64 \\
\hline
\end{tabular}
\end{table*}

The blackbody temperature of the flares varies between 26 and 40\,eV. This makes the flares from the secondary pole similar in behavior to those observed with ROSAT originating from the primary pole. Both \citet{ramsay+96} and \citet{beuermann+08} found the blackbody temperature to be variable as a function of the count-rate between 25 and 34 eV, the former study from flare to flare, the latter study even within a given flare. The higher the count rate, the hotter the emitting area. We searched for such evidence in our data, in the spectral fits to the individual flares, and within the flares, by inspecting the hardness ratio $HR = (H - S) / (H + S)$ between the count rates in a few bands that are sampling the soft component ($0.20 - 0.25$\,keV, $0.25 - 0.35$\,keV, and $0.35 - 0.50$\,keV, respectively). We could not find any systematics in any of the two hardness ratios toward higher temperatures with rising count rate (where the sum of all three bands is used). We conclude that  the temperature of the filamentary accretion area does not seem to depend on the mass or length of the infalling blobs. We cannot make statement about variability within single flares due to the lack of photons.

We find, however, a clear anti-correlation between the black-body temperature and the emitting area (see Fig.~\ref{f:flarefits}, lower left panel). We also find a weak anticorrelation between the temperature of the black-body and the bolometric luminosity, so that the dispersion of the specific luminosity becomes rather small, 
$\log\langle L_{\rm BB, bol}/f\rangle = 37.30 \pm 0.14$\,erg s$^{-1}$.

The mass of individual filaments is observed to vary by a factor of $\sim$50, the
length by a factor of $\sim$100 and the two quantities are weakly correlated. The blob
mass does not scale with the mass flow rate $\dot{m}$ but (weakly) 
with the accretion rate $\dot{M}$.

\subsubsection{X-ray and ultraviolet absorption troughs and dips}
The absolute novel feature in the 2015 data of AM Herculis is the complete extinction of soft X-rays at around phase 0.7. More precisely, it was observed to start at phase 0.6 or soon after and was observed to last 0.25 to 0.35 phase units in the softest band (Fig.~\ref{f:xlcs15}). A similar feature is observed in the UV although there it appears to be considerably shorter in phase. This suggests that the X-ray absorption is caused by two different structures in the binary. The first, starting shortly after phase 0.6 mainly affects the soft X-ray spectral regime and the UV, whereas the second dip is prominently visible up to the highest energies. This high-energy feature displays a variable central phase somewhere between phases 0.90 and 0.98. Soft X-rays are much less affected during that phase interval, and the feature is barely if at all recognized in the UV.

Contrary to past high-state observations \citep[see Sect.~3.2.3 and][]{gaensicke+95,gaensicke+98} the white dwarf does not seem to possess one well-defined heated accretion area that previously gave rise to pronounced UV-variability via fore-shortening and was used to constrain its size and temperature (see e.g.; Sect.~\ref{s:uv2005}). Given the lack of such well-behaved variability pattern we do not attempt to model the UV data in terms of a spot model. Without color information, without a clear phase-dependent variability pattern and with the likely co-existence of two or even more accretion areas the derived parameters would be degenerate and not really informative.

The simultaneity of the initial soft X-ray and UV absorption events and its duration suggests an origin in cold interbinary matter, likely an accretion curtain. The minimum countrate detected with the OM and the UVM2-filter at the bottom of the three absorption events is about 19 s$^{-1}$ which corresponds to $4.2 \times 10^{-14}$\,\flux. This is about the same flux that was estimated by \citet{gaensicke+95} originating from the accretion stream or curtain. One may thus assume that the background light from the white dwarf and accretion spot is completely blocked and the remaining emission solely due to foreground emission from the accretion stream or curtain. Under this 
assumption the absorbing column density is $\log N_{\rm H} > 21.4$\,cm$^{-2}$. 

\begin{figure}[h!]
   \resizebox{\hsize}{!}{\includegraphics[angle=-90,
   clip=]{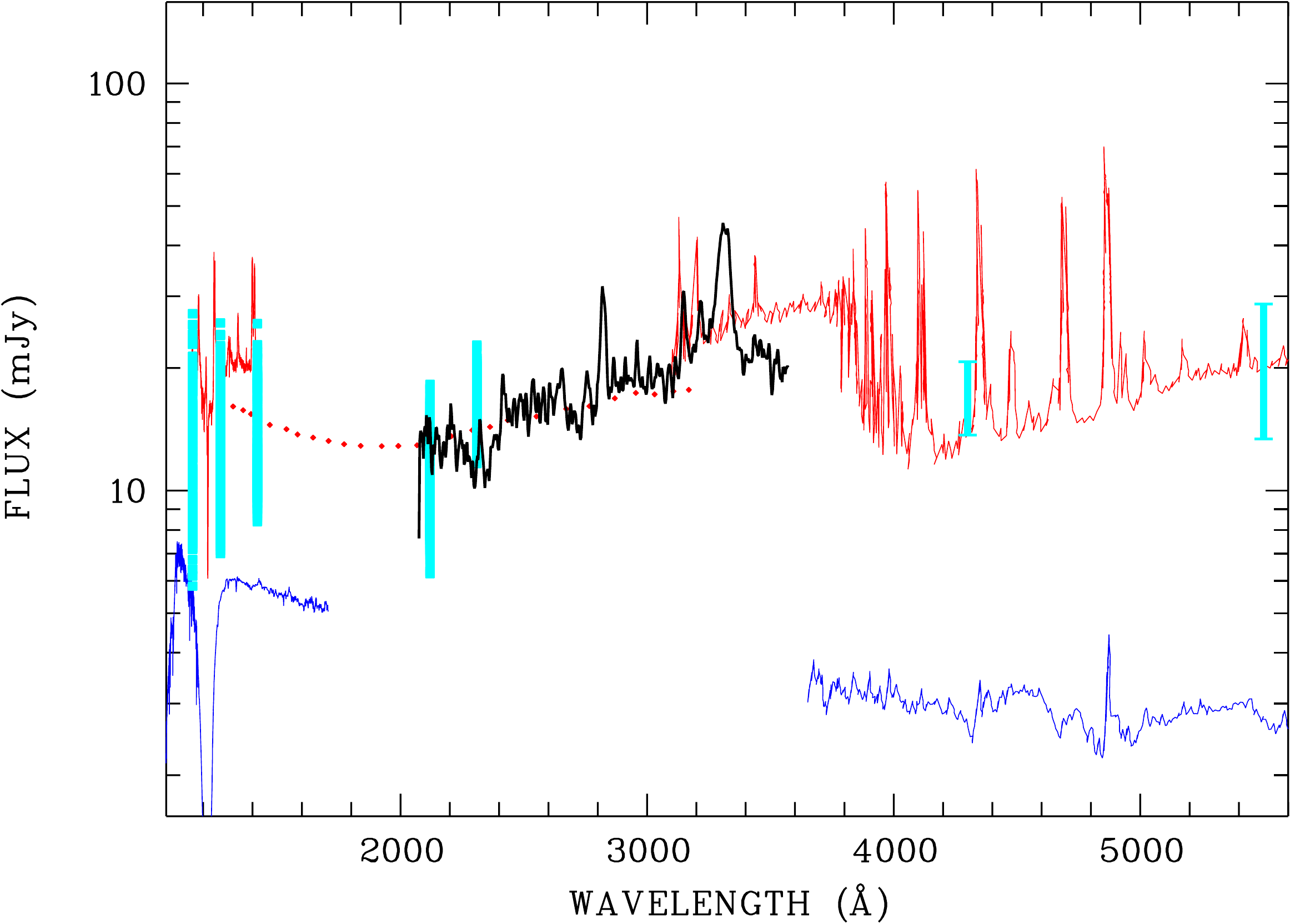}}
    \caption{UV-to-optical spectral energy distribution in low and high states. Spectra shown in  blue were obtained in low states by \citet{gaensicke+06} and \citet{schmidt+81}, those in red in high states by \citet{schachter+91} and \citet{gaensicke+98}. The OM grism spectrum shown in black is the mean of spectra \#12, 13, and 15. The vertical lines in cyan indicate phase-dependent variability from time-resolved spectroscopy and photometry with the HST/GHRS, the OM with filters UVW2 (2144\,\AA) and UVM2 (2327\,\AA), and from the AAVSO through filters $B$ and $V$. The red dotted line indicates the high-state bright-phase UV continuum measured by \citet{gaensicke+95}. 
}
    \label{f:uvsed}
\end{figure}

A similar amount of absorption was found from the spectral analysis with the UV-grism. Four spectra with the UV-grism were obtained (observations \#12, 13, 14, and 15) with spectra \#12, 13 and 15 being almost being almost identical to the high state, bright-phase IUE spectrum LWP 20132. They fall short of the IUE spectrum by 20-25\% above 2800\,\AA, likely an instrumental effect (see Fig.~\ref{f:uvsed}). About 42\% of spectrum \#14 are affected by absorption. If de-reddened with absorption coefficients by \cite{cardelli+89} the implied column density is $N_{\rm H} = 1\times 10^{21}$\,cm$^{-2}$.

We investigate the X-ray spectrum (all three EPIC cameras were used) during this phase of total soft X-ray absorption by adding a further cold absorber ({\tt TBABS}) to the plateau plus flaring blackbody spectrum. The overall normalization of the plateau spectrum and the flaring blackbody were left free to account for geometric effects.
The best fit ($\chi_\nu^2 \sim 1$ for each individually fitted dip) was achieved when the parameters of the partial covering absorber were allowed to vary freely. The best fit parameters for the partial covering absorber were $N_{\rm H} = (4.1\pm0.1) \times 10^{22}$\,cm$^{-2}$, $f = 0.93\pm0.01$ ($f$: covering fraction) and for the extra cold absorber ({\tt TBABS}) a column density of  $N_{\rm H} = (1.2\pm 0.2) \times 10^{21}$\,cm$^{-2}$, the latter value in excellent agreement with that inferred from the UV absorption dip. Hence, the picture of a highly structured absorber emerges.

\begin{figure*}[th]
\resizebox{0.45\hsize}{!}{\includegraphics[clip=]{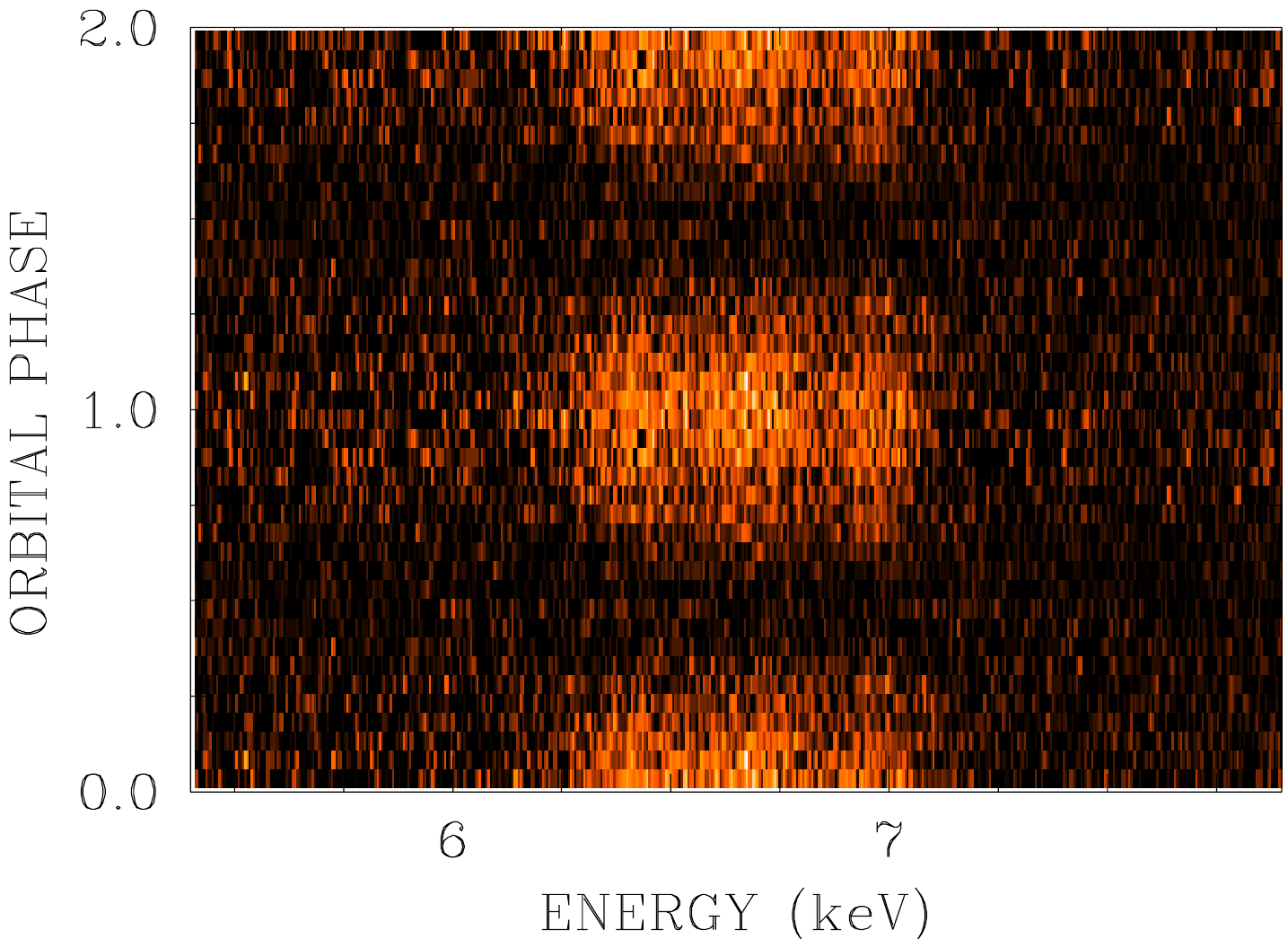}}
\resizebox{0.032\hsize}{!}{\includegraphics[clip=]{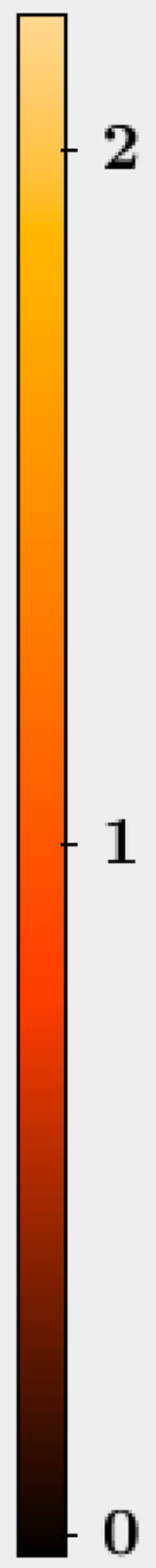}}
\hfill
\resizebox{0.45\hsize}{!}{\includegraphics[clip=]{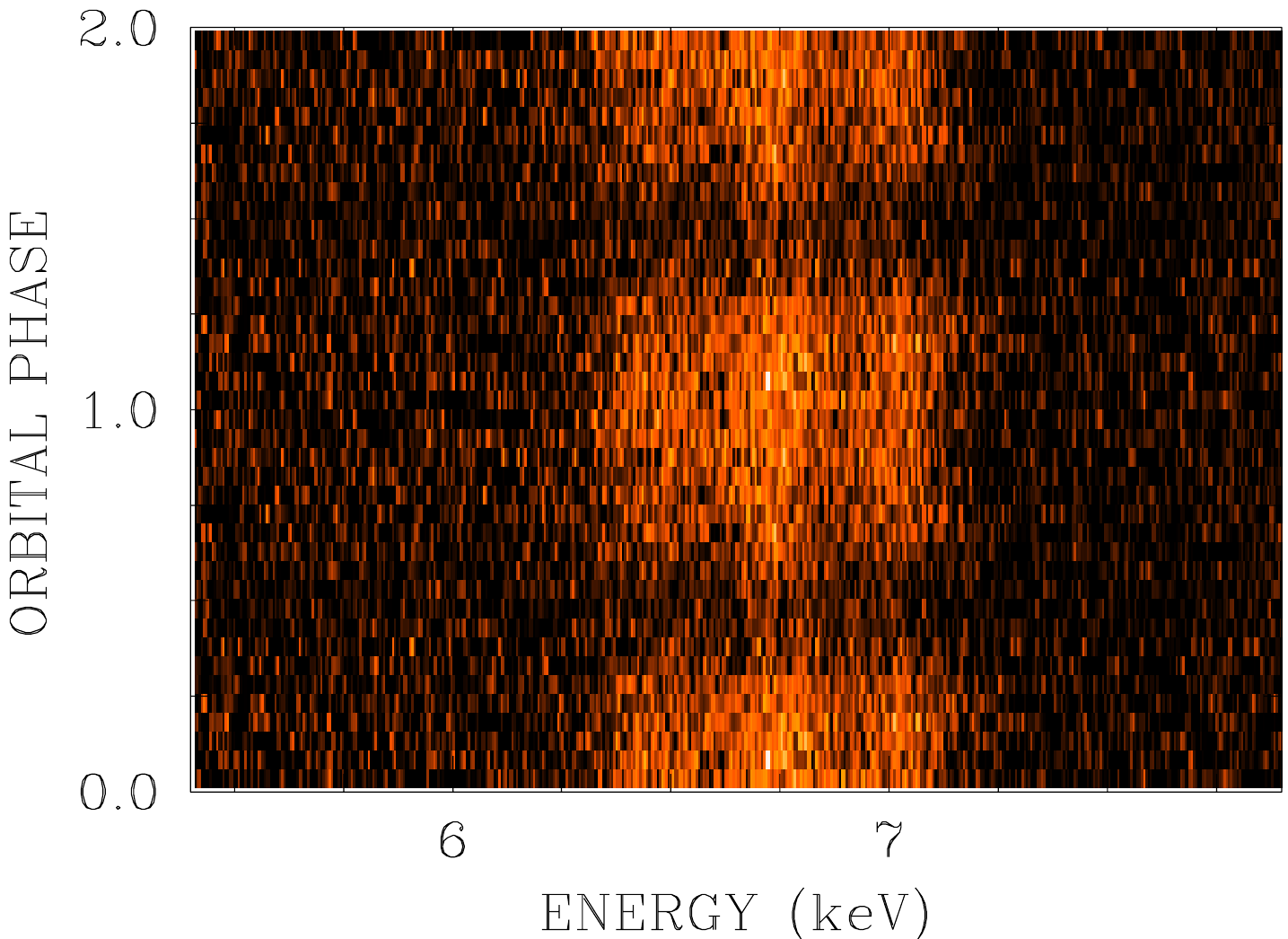}}
\resizebox{0.032\hsize}{!}{\includegraphics[clip=]{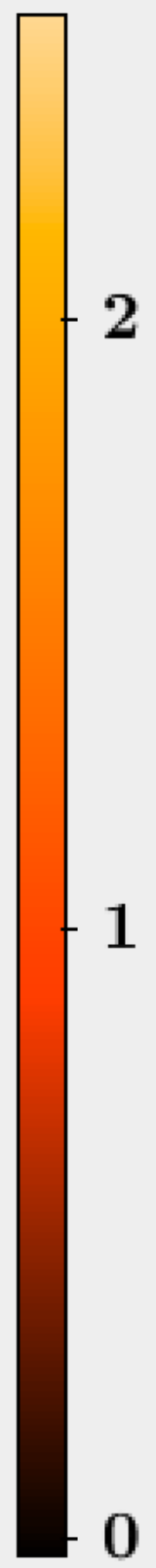}}
\\[2ex]
\hfill
\resizebox{0.49\hsize}{!}{\includegraphics[clip=]{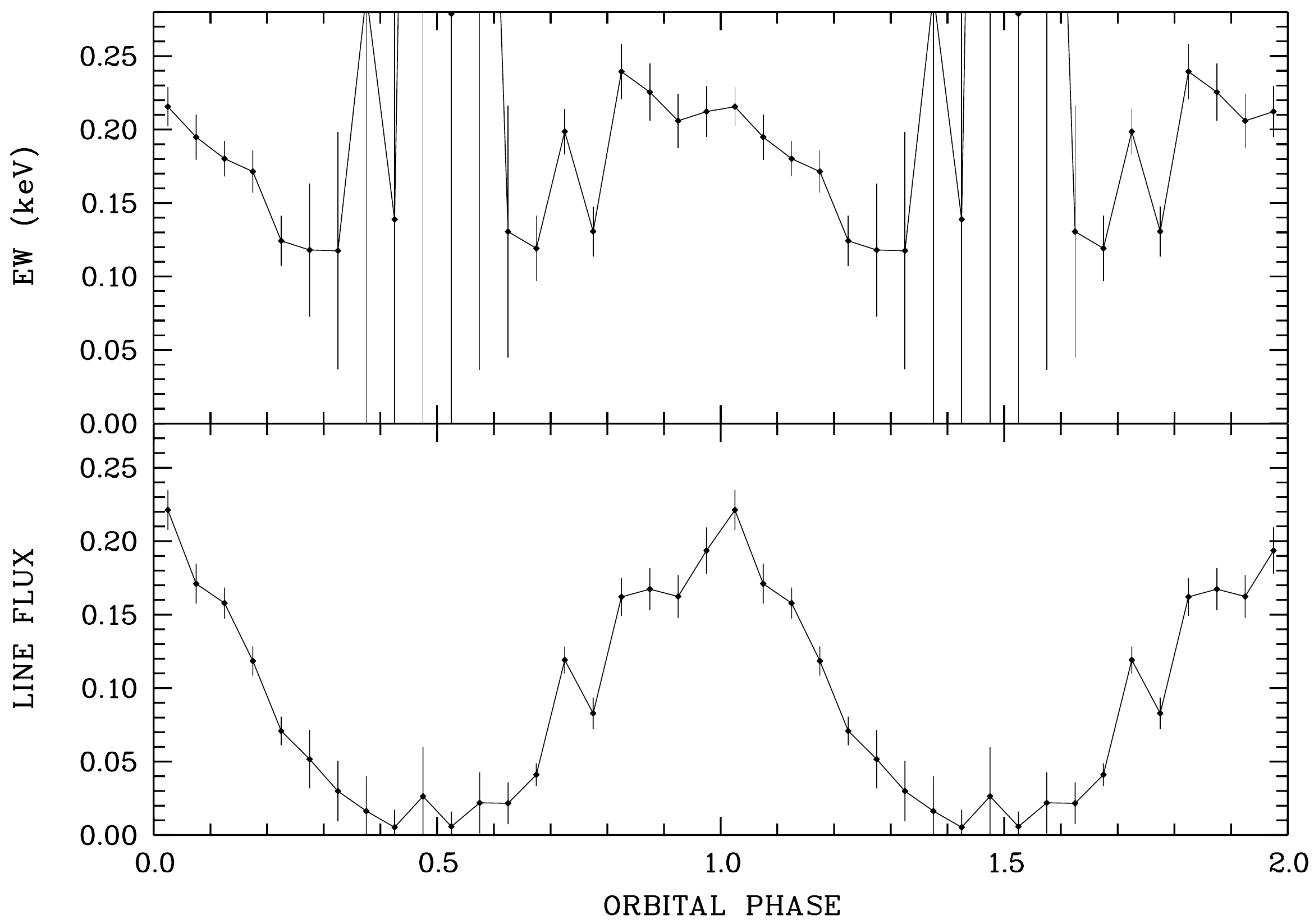}}
\hfill 
\resizebox{0.49\hsize}{!}{\includegraphics[clip=]{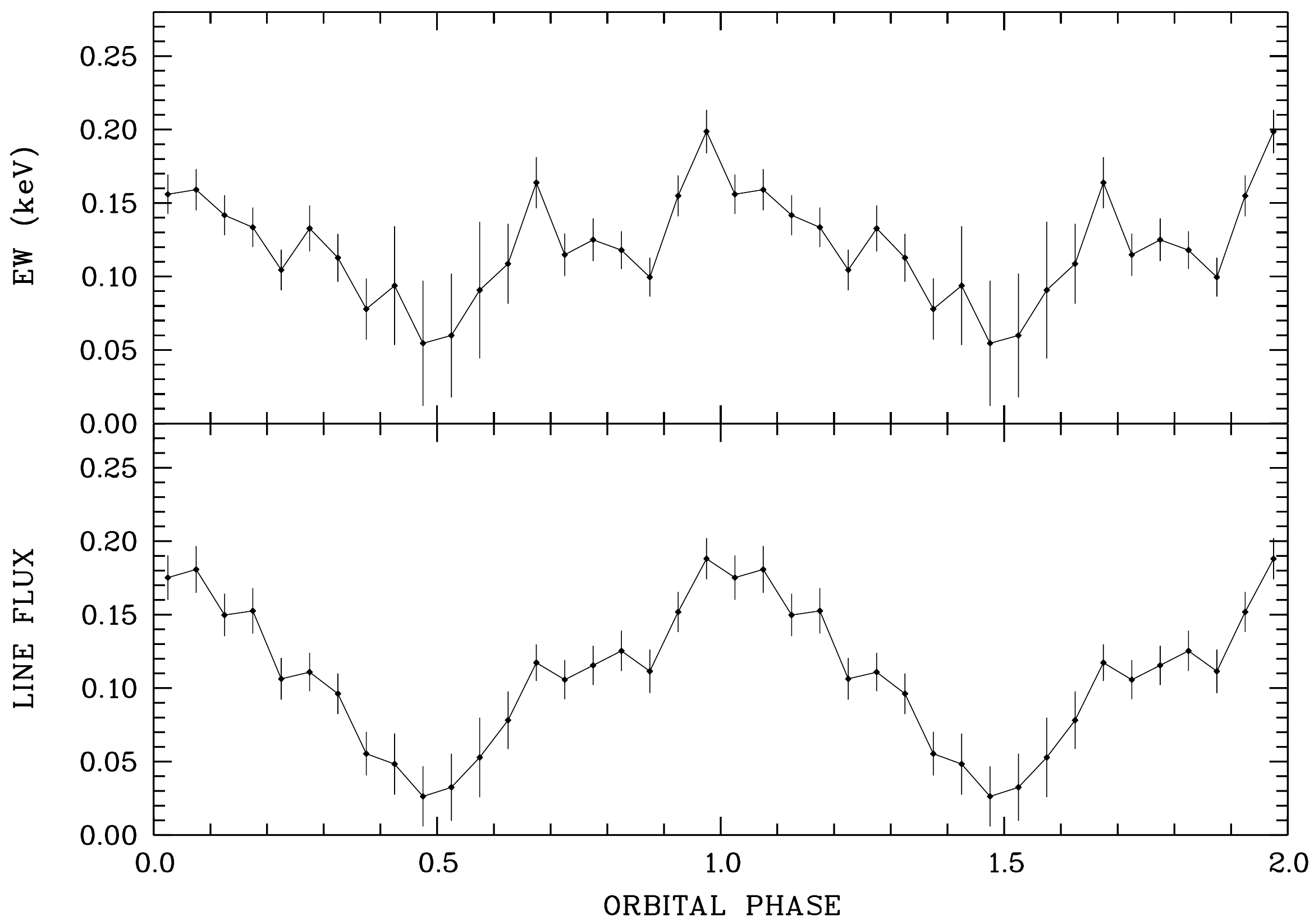}}
\caption{{\it (top)} Continuum-subtracted phase-averaged EPIC-pn spectra of \amh\ obtained in 2005 (left) and in 2015 (right) arranged as apparent trailed spectrograms. Only the region around the Fe-line complex is shown. 
The original data were arranged in 20 phase bins and energy bins of 5\,eV. The color bars indicate spectral flux in counts s$^{-1}$ keV$^{-1}$.
{\it (bottom)} Line flux and equivalent widths of the neutral Fe-line as a function of binary phase. The same data are shown twice in all panels just for better visibility of repetitive features.
\label{f:refl}}
\end{figure*}

\subsection{Phase-dependent variability of the Fe-line complex in 2005 and 2015\label{s:felines}}
A search for phase-dependent variability of the line fluxes and the equivalent widths of the iron emission lines between 6 and 7 keV was performed. We were mostly interested in the photometric variability of the neutral line thought to originate from resonant scattering, also referred to as reflection.

The EPIC-pn data of 2005 and 2015 were sorted into twenty phase bins. The continuum-normalized phase-averaged spectra were arranged as an apparent trailed spectrogram and are displayed in the first row of Fig.~\ref{f:refl}.

From a fit to the mean spectra in 2005 and 2015 for the restricted energy range between 6.0 and 7.5 keV with a bremsstrahlung component plus three Gaussian emission lines we found multiplicative factors between the line energies\footnote{These ratios were found to be different between MOS und pn in 2005 and 2015. We interpret those differences as calibration uncertainties for the two different observation modes used}. 

With the line energy ratios known, we fit the spectra of the twenty phase bins and calculated line fluxes and  equivalent widths of the 6.42\,keV line. The results are also shown in Figure \ref{f:refl} in the bottom row. Errors for the line fluxes and equivalent widths are given at the $1\sigma$ level. 

Both the line flux and the equivalent width (EW) follow at both occasions roughly the continuum variability. This locates most of the line flux origin at the white dwarf surface around the normal-accreting prime pole. The high EW around phase 0.5 in 2005 is apparent only; it is compatible with zero. 

There are nevertheless pronounced differences 
between 2005 and 2015. In 2005 the line flux was modulated by 100\% whereas in 2015 it seemed to stay finite. The maximum line flux and the maximum EW were lower in 2015 than in 2005. The line flux variability follows the pattern of the underlying continuum (not shown in the figure, but see the light curves above, e.g., in Fig.~\ref{f:lcs0515}) with the small but perhaps significant exception that the fractional variability in 2015 is about 55\% in the continuum but 85\% in the line. Would the irradiating spectrum be the same, a different maximum value of the EW would indicate a different viewing angle onto the accretion column. The absence of the self-eclipse requires a significant migration of the main accretion region. If this is sufficient to explain the change in maximum EW or if a varied irradiation spectrum also needs to be taken into account cannot be disentangled uniquely. 

Theoretically, the equivalent width of the fluorescent iron line at a given orbital phase is dependent on various parameters describing the emission and the scattering processes. The WD mass, the magnetic field  and the specific mass flow rate determine the emission properties and the height of the emitter above the WD surface, while the abundance of the ambient medium and the phase-dependent viewing geometry determine the actual strength of the fluorescent line. The viewing geometry itself depends on the orientation of the magnetic field in the accretion column, the orbital inclination and the phase angle. The abundance might also be variable as a function of distance from the accretion spot.

X-ray reflection and the behavior of the Fe K$\alpha$ lines was recently modeled by \cite{hayashi+18} as a function of the mentioned parameters. The detailed angular dependence of their post-shock accretion column (PSAC) model was not published, which only allows us to make a few qualitative statements comparing our data with their models. Firstly, an EW as large as 180 eV (as in 2015) or even larger as 200 eV (as in 2005) observed by us is not reproduced by their PSAC models. Their PSAC model for a standard mass flow rate of 1\,\mrats for an 0.7\,\msun\ WD gives an EW of 135 eV when viewed pole-on and reaches 160 eV at unrealistic high mass flow rates of $10^3$\,\mrats. Such high an EW as observed in AM Herculis requires an enhanced iron abundance (not very likely to occur in the accretion region) or contributions from a scattering medium somewhere else than the white-dwarf surface, some pre-shock material or a diluted accretion halo being excellent candidates. 

\subsection{H-like and He-like oxygen lines in the RGS spectrum \label{s:rgslines}}

\begin{figure}
\resizebox{\hsize}{!}{\includegraphics[clip=]{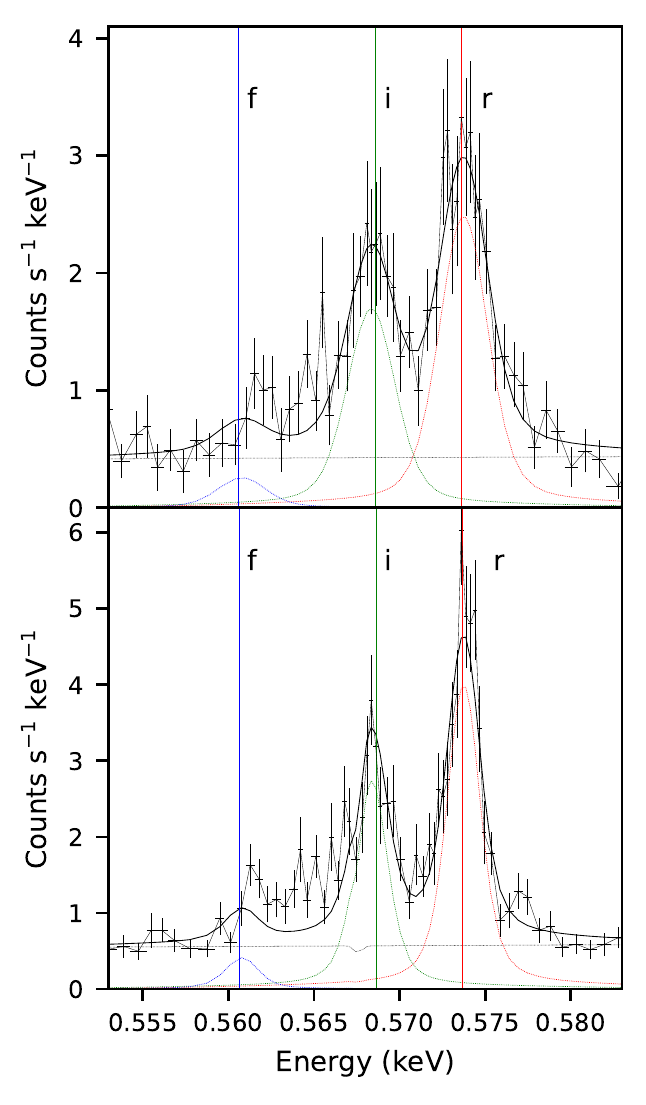}}
    \caption{Mean RGS-spectra of the 2005 (top) and 2015 (bottom) observations centered on the He-like triplet of O{\sc VII}. Vertical lines indicate the nominal line positions. Best-fit individual line components are shown in color. 
\label{f:rgslinfit}}
\end{figure}

\begin{table}
 \caption{Oxygen line fit parameters. Labels $r,i,$ and $f$ refer to the resonance, intercombination, and forbidden lines of O{\sc vii}, respectively. Line fluxes are given in units of $10^{-13}$\,erg cm$^{-2}$ s$^{-1}$. The first error term for the $i$ and $f$ lines indicates statistical, the second a systematic error which takes the excess photons between the lines into account. The last row gives the line ratio between the two species.
}
 \begin{tabular}{l|cc}
            & 2005        & 2015                                  \\ \hline
O{\sc vii} $E_r$          & $573.81 \pm 0.11$\,eV    & $573.77 \pm 0.06$\,eV  \\
O{\sc vii} $\sigma$       & $1.1 \pm0.1$\,eV         & $0.54 \pm 0.10   $\,eV  \\
flux $r$                  & $2.64 \pm 0.18$          & $3.45 \pm 0.18   $ \\
flux $i$                  & $1.81 \pm 0.16 \pm 0.13$ & $2.29 \pm 0.16 \pm 0.14 $ \\
flux $f$                  & $0.28 \pm 0.12 \pm 0.03$ & $0.35 \pm 0.10 \pm 0.01$ \\
O{\sc viii} $E$           & $653.16\pm 0.15$\,eV     & $653.09 \pm 0.13 $\,eV \\
O{\sc viii} $\sigma$      & $1.0  \pm 0.2$\,eV       & $1.1  \pm 0.2    $\,eV \\
O{\sc viii} flux          & $1.61 \pm 0.10$          & $2.16 \pm 0.11   $ \\
$G = (f+i)/r$             & $0.79 \pm 0.17$          & $0.76 \pm 0.12   $ \\
$R = f/i$                 & $0.15 \pm 0.08$          & $0.15 \pm 0.05   $ \\
H/He ratio &                $0.34 \pm 0.05$          & $0.35 \pm 0.04   $ \\
\end{tabular}
\label{t:rgslinfit}
\end{table}

In this subsection we wish to determine the location of the origin of the lines found in the RGS data with temperature and density diagnostics allowed by a line ratio analysis. Only the H-like O{\sc viii} line and the Helium-like O{\sc vii} triplet are detected with sufficient signal in the RGS data. Previous brief accounts  on those lines were given by \cite{burwitz+02} and \cite{girish+07} based on Chandra LETG and HETG observations, respectively. While \cite{burwitz+02} report significantly broadened O{\sc vii}, which they ascribe to phase-dependent Doppler shifts that lead to broadening in the phase-averaged spectrum, \cite{girish+07} used line ratios to estimate electron temperatures and densities as 2 MK and $>2\times10^{12}$\,cm$^{-3}$. 

The electron density at the top of a standard accretion column with a specific mass flow (accretion) rate of $1$\,g\,cm$^{-2}$\,s$^{-1}$ and with solar composition is expected to be of order $7.5 \times 10^{15}$\,cm$^{-3}$, and more than an order of magnitude higher at its base and we test whether this estimate is consistent with the density-sensitive He-like triplet lines of OVII.

Ratios of Hydrogen-like to Helium-like lines and the line ratios of He-like triplets are widely used for diagnostics of a collisionally ionized and hybrid plasmas \citep[see][ and references to this paper]{porquet_dubau00}, the latter via the ratios of the forbidden $(f)$, the intercombination $(i)$ and the resonance lines $(r)$.  Analytical relations between those depend on the electron density $n_e$ and the electron temperature $T_e$ are $G(T_e) = (f+i)/r$ and $R(n_e) = f /i$ \citep{gabriel_jordan69}. 

For both observational campaigns the RGS data from both instruments were combined (and for 2005 all four revolutions, except OBSID0305240301, which did not contain RGS data) using the SAS task {\tt rgscombine} and binned with a minimum of 20 counts per spectral bin. To derive line fluxes the resulting combined spectra were fitted with a bremsstrahlung component with fixed temperature at 15 keV to provide a background and Gaussians superposed. For the Helium-like triplet three Gaussians were used for the recombination line at 21.6015\,\AA, the intercombination line at 21.8036\,\AA, and the forbidden line at 22.0977\,\AA. The energies of the lines were fixed to each other according to their known ratios (taken from AtomDB) and their widths assumed to be equal. There was some uncertainty about the line strength of both the intercombination and the forbidden lines due to an excess of photons between the two lines. Thus the triple Gaussian fits were not found to be fully satisfactory (see Fig.~\ref{f:rgslinfit}) and some doubt remained about the mere existence of the forbidden line.

We tested for its existence by evaluating the changes of the $\chi^2$ statistics when the line flux was finite or set to zero. Following \cite{wall_jenkins12} (their equation 6.11) we find probabilities of 97.3\% and 99.999\% that the improvement in $\chi^2$ when the line is included is not due to a statistical fluctuation but due to an excess of flux at the predefined position of the forbidden line. We, therefore, proceed under the assumption that the forbidden line is present.

The line parameters as determined from the triple Gaussian spectral fits are listed in Tab.~\ref{t:rgslinfit} with their statistical errors. 
To compute $G$ and $R$ ratios an additional 10\% systematic uncertainty of the fluxes of the $i$ and $f$ line components were taken into account. This additional uncertainty was quantified by adding a further unresolved Gaussian between the forbidden and the intercombination lines at 565.0\,eV and studying the impact on the flux of the adjacent lines.
To derive the parameters of the O VIII line at 18.97\,\AA\ neither the line energy nor the widths were predetermined and fixed.

Using our fits to the spectra obtained in 2005 and in 2015 we consistently derive an H/He line ratio of 0.35, $R=0.15$ and $G=0.76 - 0.79$ (see Tab.~\ref{t:rgslinfit}). The Chandra MEG-data analyzed by \cite{girish+07} yielded $R < 0.48$ and $G<0.76$, both values are in agreement with our data and analysis within the errors, although the RGS data are more constraining regarding the value of $R$. 

The measured line ratio of H-like to He-like lines implies a temperature of 0.2 keV for a plasma in CIE ionization balance \citep{mewe+85}. $G$ and $R$-line ratios were computed for different ionization conditions by \cite{porquet_dubau00}. The measured $G$ ratio consistently implies an electron temperature of $2.5^{+1.0}_{-0.6}$\,MK ($165 \dots 300$ \,eV) while the derived density, taken the measured $R$ ratio at face value, becomes $n_e \simeq 7 \times 10^{11}$\,cm$^{-3}$. 

The implied temperatures from both the Helium-like triplet and the H/He-like line ratio is compatible with the lowest T of the multi-temperature APEC fit to the overall spectrum (Sect.~\ref{s:2015bri}). Hence, based on the temperature estimate the likely line formation region lies at the bottom of the accretion column. However, according to published shock models \citep[c.f.][ see the estimate above]{imamura_durisen83, fischer_beuermann01} the density at the bottom of the column is several orders of magnitude higher than derived from the line ratios. Hence, our measured line ratio $R$ is not compatible with shock models but is likely to be understood in the framework of a structured accretion region with a coronal plasma at a variety of densities. Analyzing XMM-Newton/RGS and CHANDRA/HETG spectra of the intermediate polar AE Aqr, \cite{itoh+06} and \cite{mauche09} derived similar densities in the line formation region of this object. This led \cite{itoh+06} to conclude that the X-ray emitting plasma in AE Aqr cannot be a product of mass accretion onto the white dwarf. 

For AM Herculis such a far reaching conclusion cannot be drawn, since X-ray emission in polars is proven to originate from small accretion-heated spots through eclipse studies \citep[for example,][]{schwope+01} but the very low density of the line formation region is puzzling. 
The discrepancy between accretion column models and observations of MCVs might even be larger because the background UV radiation field from the white dwarf and the accretion hot spot will lead to de-excitation of the upper level of the forbidden line. Hence the density is overestimated when leaving out the effect of the radiation field \citep{ness+01, mauche02}. \cite{mehdipour+15} have shown for a photoionized plasma that the value of $R$ of the O{\sc VII} triplet may be changed by a large amount if absorption (in particular of the intercombination line through Li-like oxygen) is taken into account. This effect gives a bias of the density diagnostics in the opposite, hence desired, direction than photo-dexcitation. However, since the X-ray spectrum is otherwise well explained by a collisionally ionized plasma we do not expect this effect to have a fundamental impact on the line ratio analysis. There is a clear desire to better understand the line formation scenarios in a structured accretion region with a interplay between collisions and photoionization.

The measured line positions of both species considered here give some weak evidence at the $1-2 \sigma$ level for a positive radial velocity of order 100 -- 200\,\kmps\ for 2005 and at $3\sigma$ for 2015. Such a velocity would be consistent with an origin from the bottom of an accretion column. We also searched for a relative line shift between orbital maximum and the phases when the accretion column is at the limb but those experiments yielded weak constrains only with $\Delta v < 400$\,\kmps. Although our radial velocities have large error bars  we cannot confirm any significant radial velocity variations of the oxygen lines of $\simeq 750$\,\kmps\ reported by \citet{burwitz+02} on the basis of Chandra/LETG spectra obtained back in the year 2000 (September 30) during  another high state of AM Herculis. A re-analysis of the Chandra/LETG spectra performed by us following standard recipes did not reveal a significant line shift either. It is evident to us, that reliable measurements of line shifts require data with much higher S/N than currently available through grating spectroscopy obtained with XMM-Newton and Chandra and typical average exposure times.
 
The measured line widths is slightly larger than the instrumental resolution. Formally the measured line widths of O{\sc vii} were slightly discrepant but the differences stay below 3$\sigma$. The finite line width is difficult to interpret, since projected orbital and settling velocities in the column contribute to thermal broadening and cannot be disentangled. 

\begin{figure}
\resizebox{0.915\hsize}{!}{\includegraphics[clip=]{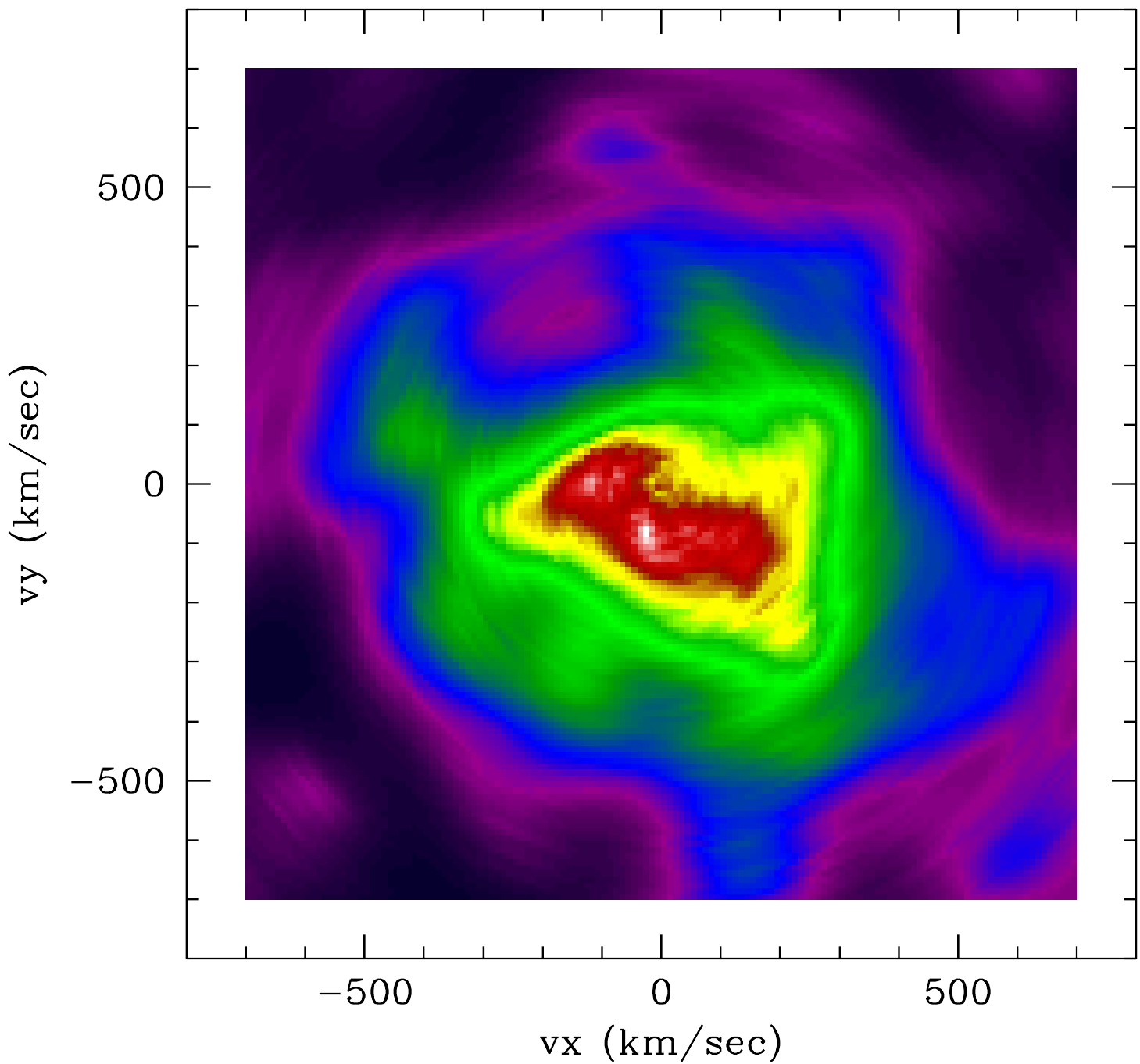}}
\hfill
\resizebox{0.073\hsize}{!}{\includegraphics[clip=]{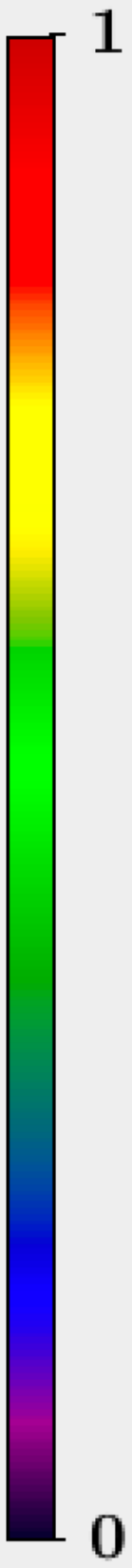}}
\caption{Doppler map of H$\alpha$ based on spectra obtained on April 19, 2015, with CAFE. The values in the map were normalized to a range between 0 and 1.
\label{f:dop}}
\end{figure}

\subsection{Tomographic analysis of the 2015 CAFE spectra}
The relative faintness of our target (for a 2.2m aperture and a high-resolution spectrograph) together with our wish to obtain phase-resolved data led to noisy individual spectra. A continuum was difficult to recognize and only the most prominent lines like H$\gamma$, He{\sc II}\,4686, H$\beta$, He{\sc I}\,5875 could be recognized in the mean spectrum or traced through the binary cycle (H$\alpha$). Despite the rather low signal-to-noise ratio of our spectra it became clear that the narrow emission line (NEL) that was prominent in a past high accretion state \citep{staude+04} was absent in the new data. The NEL, which is a prominent feature in many polars \citep[see, e.g., the review by ][]{marsh_schwope16}, originates from the UVX-irradiated hemisphere of the donor star and was used to trace the donor in the binary system and constrain the mass ratio.

The H$\alpha$ Doppler map shown in Fig.~\ref{f:dop} underlines the impression from visual inspection of the trailed spectrogram. There is no enhanced emission from the location of the donor star. Otherwise the map shows relatively little structure, emission is found in all quadrants, with most of the emission at negative $y$-velocities and with both negative and positive $x$-velocities. A negative $v_x$ indicates matter streaming from the donor to the white dwarf, a positive $v_x$ in the opposite direction. The optical and the high-energy observations with XMM-Newton and NuSTAR were taken two weeks apart. It is tempting to assume that \amh\ was still in the reversed mode of accretion when the optical data were taken. If so, the map can be interpreted by assuming that the accretion curtain which gives rise to the prominent soft X-ray absorption prior to phase zero completely shields the donor star thus explaining the absence of the NEL. Streaming matter feeding all the  accretion regions at near and far sides of the white dwarf as seen from the donor star leads to broad emission lines at various Doppler velocities, which cannot be disentangled and attributed to just one of the accretion streams or curtains. In that regard the new map is reminiscent to those of the asynchronous polars BY~Cam and V1432~Aql \citep[][their figure 11.14]{marsh_schwope16}.

\begin{table*}
\caption{Summary of the XMM slew observations of \amh. Hardness ratios and fluxes (from 0.2 to 10.0\,keV in units of $10^{-12}$\,erg\,s$^{-1}$\,cm$^{-2}$) are given for the slews and for the pointed observations at the same orbital phase.}
\begin{tabular}{lllllll}
slew ID    & Date        & $\phi_\text{orb}$ & HR/flux  & HR/flux  & HR/flux  & comment \\
           &             &                   & Slew     & 2005     & 2015     &          \\ \hline
9170900002 & April 09, 2009 & 0.84              & -0.4/50  & -0.7/215 & +0.2/180 & Normal mode, low/intermediate state \\
9173100003 & May 23, 2009 & 0.72              & +0.3/140 & -0.5/180 & +0.1/140 & Reversed mode \\
9189300004 & April 11, 2010 & 0.605             & +0.4/40  & -0.7/43  & -0.6/260 & Previously uncharacterized faint hard state \\
9195400002 & August 10, 2010 & 0.72              & -0.45/75 & -0.5/180 & +0.1/140 & Normal mode \\
\end{tabular}
\label{t:slews}
\end{table*}

\subsection{XMM-Newton slews}
AM~Her was observed four times by XMM-Newton during slews
\citep{Saxton+08}. All slews were taken in full frame or extended full frame
mode, with the medium filter. Though the effective on-target times are short
($\sim 10-15$\,s) enough photons may be detected to distinguish between the
normal and reversed accretion modes by hardness ratios and rough flux
estimates. 

We downloaded the slews and reduced them with the {\tt epproc} and {\tt eslewchain}
tasks, then barycenter-corrected the resulting event lists to calculate the
orbital phases during which the slews occurred. Fluxes in the
0.2-12\,keV band, and count rates in the hard and soft bands, are taken from
the slew catalog. The results are summarized in Table \ref{t:slews}.

We cannot compare hardness ratios of the former to the latter because different filters were used. Instead, we extracted spectra for the corresponding phase
intervals for the PN instrument and produced spectral models that closely fit
the continuum. These models were used to make fake spectra with 15\,s exposure
time and an ancillary response file for the medium filter. Photon counts in
the hard and soft energy bands are thus directly comparable to count rates in
the slews. 

The first observation occurred just before the
beginning of a long high state (see Fig.~\ref{f:aavso_99_16}, MJD 54930) and during a period of
variability between low and intermediate states. Its orbital phase was near
the peak visibility of the primary accreting pole. The soft hardness ratio
suggests that the accretion more resembled the normal rather than the reversed
mode (cf.~Fig.~\ref{f:lcs0515}). We suspect, therefore, that this slew
occurred during a brief intermediate state and that this exhibited the normal
accretion mode.

The second slew observation was just after the beginning of the same high
state. Its hardness ratio and flux are similar to the reversed state 
of 2015, and was therefore likely observed during similar behavior. This result
suggests that switches between normal and reversed modes are not very rare.

Occurring slightly before a brief return to the low accretion state, the third
slew observation has a very hard spectrum. Both of the pointed observations
around this time had much softer spectra at this phase. It would have
coincided with the intense flaring if in the reversed accretion mode. The
flares persist for over a minute, and the gap between flares are of similar
duration (see section \ref{s:flaring}), but the slews represent a snapshot of only
10-15\,s. We therefore repeated the 2015 fake slews twice, once for the gap
between flares and once during a flare. The flare is much
more luminous and softer, but even the between-flare spectrum does not
resemble the low flux and hard spectrum observed in the slew from April 11, 2010. We
therefore cannot identify which accretion state this slew corresponds to. It
hints at the existence of other, previously uncharacterised, modes of accretion.

The final slew observation happened near the end of a rise toward high state after a short low state (MJD 55418), with orbital phase the same to that of the second slew. It has a hardness ratio similar to the normal (2005) mode of accretion but at a
significantly lower flux.

\section{Discussion}
In this paper we have analysed comprehensive X-ray observations obtained with \xmmn\ and NuSTAR at X-ray and ultraviolet wavelengths and contemporaneous optical observations of the prototypical polar \amh. These were obtained in high accretion states in 2005 and 2015. In 2005 \amh\ was encountered in its regular mode of accretion, that means it had one X-ray bright accreting pole that was self-eclipsed for about 0.16 phase units. The spectrum could be described with a soft (blackbody-like) and a hard multi-temperature thermal component. No pronounced soft flaring was observed and the soft X-ray excess was rather moderate, $F_{\rm bb,bol}/F_{\rm X,bol}=3.4-10$ correlated with the X-ray brightness, hence accretion could be seemingly described with a standard accretion column model. 
The spectral parameters reported here are broadly consistent with those reported for previous high-state observations obtained with ROSAT, EUVE, RXTE, and Chandra/LETG \citep{beuermann+08, beuermann+12, paerels+96, christian00}. The total flux in the soft component reported by \citet{beuermann+08} is higher than ours for 2005 through the inclusion of the far-UV component derived by \cite{greeley+99}, their bolometric flux was $F_{\rm bb} = 4.5 \pm 1.5 \times 10^{-9}$\,\fint, with an error that was assumed to encompass the spread between individual high states according to \cite{hessman+00}. The soft X-ray flux varied between the four XMM-revolutions between $(0.7 - 3.3)\times10^{-9}$\,\fint, which only includes our observed fluxes and does not make an attempt to correct for the unobserved EUV flux. 

During the self-eclipse of the main accretion region observed in 2005 some basal flux was remaining and was sufficiently bright to extract an X-ray spectrum. Several components needed for a synthesis, two thermal, a soft blackbody-like component and a possible Fe emission line. In particular the presence of the soft component let us believe, that accretion at a second region is responsible for the remaining flux in the self eclipse. The line parameters are poorly constrained only but seem to imply a rather large equivalent width of about 0.7\,keV, larger than compatible with direct emission from an accretion plasma or reflection. This may hint to scattering of X-rays into the observer's line-of-sight.

A spectrum of the minimum phase was also extracted by \cite{ishida+97}.
They found an Fe-line EW of 1-2 keV and concluded that such high a value is
possible only for matter of solar abundance and the direct emission being
completely blocked and the emission above the iron emission line energy range
is scattered emission only. If this were true for our 2005 observation as well, we were in need to assume disjunct soft and hard emission regions, because the mere presence of the soft component requires a direct view. The mass accretion rate at the second pole is $\dot{M} = 5.5 \times 10^{-13}$\,\mrat, about a hundred times lower than at the primary pole.

In 2015 \amh\, was found in its reversed state of accretion with (at least) two bright accretion regions. One, still called the primary region was located somehow close to the same primary region as in the regular mode of accretion and displayed similar spectral properties. Its bright phase was centered at around phase 0.0. The other region was dominated by soft flares and was bright around phase 0.5. The phasing and spectral properties are similar as for the initial discovery of this accretion mode with EXOSAT some 30 years ago \cite{heise+85}. The novel feature reported here for the first time and observed in four consecutive binary cycles was the complete extinction of soft X-rays for more than 20\% of the binary cycle. The onset of this feature which must be caused by an extended absorbing structure like a curtain varied by $\sim$0.08 phase units between 0.63 and 0.71. A second narrower absorption feature, an absorption dip, occurred between phase 0.9 and 0.95 and was observed as a distinct feature in the highest energy bands covered by XMM-Newton while it seemed to be intermingled with absorption in the curtain at low energies.

The soft pole observed in 2015 emits a higher X-ray flux than the nominal prime pole. The summed mass accretion rate of the two accretion regions was found to be of same order than that of the prime pole in the normal mode of accretion observed in 2005 (compare the results compiled in Tabs.~\ref{t:fitbri05} and \ref{t:plateau_fit}). Hence, two-pole accretion in the reversed mode does not seem to be triggered by an enhanced overall accretion rate in a way that there is some overshooting material that is accreted at a second pole. The physics and the geometry must be more complex. Perhaps at times matter is threaded closer to $L_{\rm 1}$ and then guided to remote places in the Roche lobe of the white dwarf. Also accretion along directly coupled field lines between both stars might be possible.

A self-eclipse of this main region was observed at optical and X-ray wavelengths. The duration of the X-ray self-eclipse during our observations in 2005 was $\Delta\phi_{\rm X,2005} \simeq 0.16$ but a bit difficult to measure unequivocally. Circulation polarization sign reversals, that also trace the self-eclipse, were typically observed to last a little longer, $\Delta\phi_{\rm circ} = 0.20, 0.22,$ or 0.27 according to various authors \citep{priedhorsky+78, bailey+84, piirola88,tapia77}. The difference can be understood by realizing that the polarization signal indicates the center of the accretion region while the X-ray photometric signal is determined by the edges of the accretion region. The orbital inclination of \amh\ is not well known and may lie between 35\degr\ and 50\degr. The duration of the circular polarization sign reversal requires $i+\delta=97\degr - 102\degr$, only weakly dependent on the assumed value of $i$ ($\delta$ being the co-latitude of the field lines in the accretion region). The co-latitude of the accretion spot, which we refer to as $\psi$, might be different, and its value is definitely smaller than that of $\delta$ in a dipolar field geometry.

\begin{figure*}[t]
\resizebox{\hsize}{!}{\includegraphics[clip=]{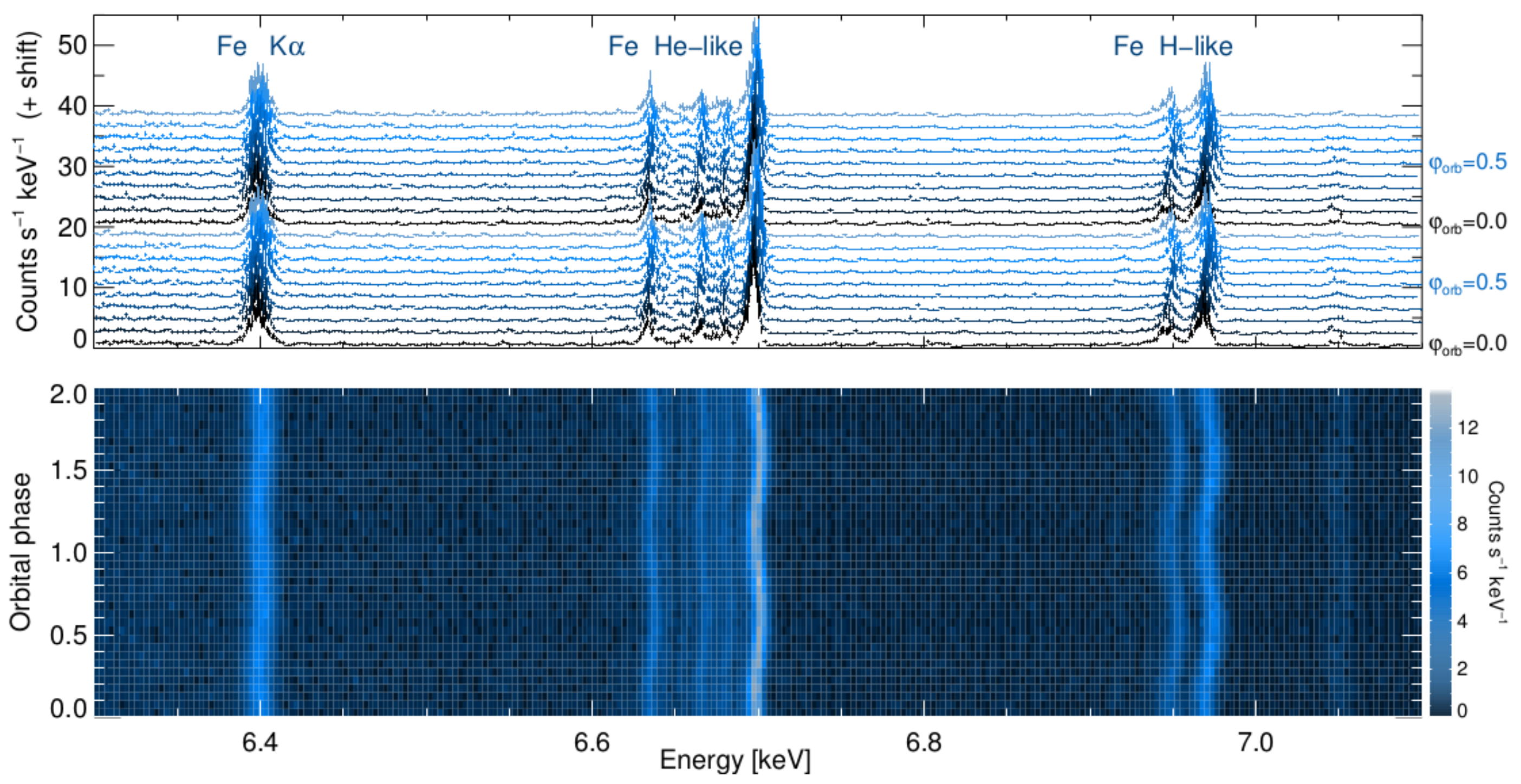}}
\caption{Simulation of a 100\,ks observation of AM Herculis with the ATHENA X-IFU neglecting the self-eclipse with a phase resolution of 0.1. For better visibility of repeated features the same data are shown twice. In the upper panel linear flux limits were chosen, in the lower panel a logarithmic scale was used to highlight subtle differences.
\label{f:athena}}
\end{figure*}

During our X-ray observations in 2015 a self-eclipse of the normal accreting pole was no longer observed, similar to the EXOSAT observation some 30 years ago. \cite{heise+85} explained the absence of the self-eclipse in the EXOSAT ME data by hard X-ray emission from the secondary pole. While this is possible such an explanation is not satisfactory because no increase in the hard X-ray emission is observed at the time when the soft flaring sets in. This is true for both the 1983 EXOSAT and the 2015 XMM-Newton observation.

An alternative explanation involves migration of the spot toward smaller $\psi$, which means that it moves farther away from the orbital plane, by some 5\degr\ to 10\degr. This explanation also has its difficulties if one assumes that this region is fed by a stream which initially follows a ballistic trajectory in the orbital plane and is then funneled to the accretion region in a dipolar geometry. The extent of a ballistic stream can be constrained by the extent of gaseous emission in Doppler maps. Given the known period and estimated stellar masses of \amh\ the maximum velocity $\abs{v_x}$ is 500\,\kmps corresponds to an azimuthal extent of the stream of less than 20\degr\ as seen from the white dwarf. The emerging accretion arc, the foot-points of all dipolar field lines which connect the white dwarf with the ballistic stream between $L_1$ (azimuth 0\degr) and the maximum extent (azimuth 20\degr)\ 
was found to vary in co-latitude by less than two degrees. Hence, a spot migration due to late or early coupling as a response to a variable mass accretion rate has too little an effect on the co-latitude of the spot to explain the vanishing self-eclipse. This leaves a non-dipolar geometry as possible remedy for the observed larger shift of the accretion region. A non-dipolar field geometry and morphology is often invoked in studies of polars to explain observations that do not conform with a simple centered dipole model, for instance to explain the complex phase-resolved photospheric Zeeman patterns \citep[recently by ][]{beuermann+20}. Here we find evidence from the variable self-eclipse in the X-ray light curves.

Even more puzzling is the geometry of the soft pole and the absorbing curtain. We firstly discuss the location of the soft pole. As described in Sect.~\ref{s:xlcs} the emission is centered on phase $\sim$0.53 and lasts less than 0.5 phase units, hence is originating from the lower ("southern") hemisphere and from that side of the white dwarf pointing away from the donor star. Its exact location is difficult to determine because we were unable to measure the end of the flaring phase precisely due to complete soft X-ray absorption. However, the general picture does not depend on these subtleties. The flaring phase of the EXOSAT observation also has a duration of less than 0.5 in phase and likely the same center. 
The remaining question is, how most of the matter manages to run around the white dwarf to be accreted at the far side (as seen from $L_1$) without a significant increase in the mass accretion rate. Such accretion regions at odd places were described in several polars and pre-polars \citep[see][ for UZ For, VV Pup, and WX LMi, respectively]{schwope+90,schwope_beuermann97,vogel+11}, although in those mentioned systems the far pole was always the faint one. Answering this question by detailed MHD simulations therefore is of general interest for accretion physics in a strongly magnetic environment.

The absorption curtain is a far or distant structure as seen from the white dwarf, i.e.~likely several $R_{\rm WD}$ away, because it absorbs radiation from both poles at the same time. The energy-resolved light curves and our spectral analysis reveal a highly structured absorbing curtain with a minimum column density of $N_{\rm H}$ of $10^{21}$\,cm$^{-2}$. Since it is a far structure with an onset observed as early as phase $\phi = 0.63$, i.e.~it is in the hemisphere behind the white dwarf as seen from the donor star. One thus could naively associate the matter in this absorbing structure with accreted matter that feeds the second, soft pole. This picture will not work because the absorbing matter is above while the soft pole is below the orbital plane. Also, there was no such absorbing structure in the reversed mode observed by EXOSAT \citep[c.f.~][]{heise+85}, but the soft pole was nevertheless fed. It could be that the accreted matter from this curtain feeds an extra region in the upper hemisphere whose radiation fills the self-eclipse of the prime pole.

The detection of an absorbing curtain and the phase-scattered absorption dips that were prominently observed in the hard X-ray light curves argue for an inclination at the higher end of the range currently discussed, which lies between 35\degr\ and 60\degr, according to \cite{brainerd_lamb85} and our analysis above. In a dipolar field geometry -- the field will be dipolar at several white dwarf radii -- matter following the field lines cannot be lifted sufficiently high above the plane to give occultation events at low inclination.  The combination of the relevant angles $i$ and $\beta$ ($\beta$: co-latitude of the accretion spot) requires $i>\beta$ to give absorption events and $i+\beta \simeq 105\degr$ to reflect the duration of the self-eclipse.

The UV data obtained with the OM were strikingly dissimilar between 2005 and 2015. The modulations  of the light curve observed through the UVW2 filter ($\lambda_{\rm eff} = 2144$\,\AA) in the regular accretion mode in 2005 could be understood in terms of foreshortening and self-eclipse of an accretion-heated spot. The flickering and to zeroth order flat light curve through the UVM2 filter ($\lambda_{\rm eff} = 2327$\,\AA) in the two- or multiple-pole mode of accretion lacks any clear sign of fore-shortening. In our modeling of the UV emission in 2005 a stream contribution of $3.9 \times 10^{-14}$\,\flux\ was used. The stream/curtain was brighter by about 20\% in 2015, as determined through the remaining flux in the absorption dips. The UV-flux at phase 0.5, i.e.~at the nominal phase of the self-eclipse of the prime pole, was about 70\% higher in 2015 than in 2005. Hence, a prominent extra source of UV-light was present in 2015 which was not self-eclipsed. This could be the white dwarf heated through accretion along the extended accretion curtain. 

Simultaneous NuSTAR observations obtained in 2015 allowed us to uncover without any doubt the long sought-for reflection component and to perform a phase-resolved analysis which suggests scattering at the white-dwarf surface. This picture is supported by the phase-dependent variability of the neutral Fe-line at 6.4\,keV (resonant scattering line) which could be traced through the orbital cycle for the first time. Both the variability and the peak value of its equivalent width are in the range expected for scattering at the white-dwarf surface. 

The oxygen lines studied with the RGS are formed in an environment with $kT$ corresponding to about 200 eV and at low velocity which hint to the bottom of the accretion column. The weakness of the forbidden line of He-like oxygen is usually interpreted as being formed in the high-density limit. Photodeexcitation through UV photons from the white dwarf and the accretion region leads to an yet to be quantified overestimated electron density, which on the other hand is several order of magnitudes below that expected at the bottom of the accretion column anyhow. A more detailed understanding of the line formation conditions requires the inclusion of more He-like species that are sensitive for other density ranges and better resolved data for radial velocity studies. Such studies will be feasible with the X-IFU onboard ATHENA (see below).

\section{Conclusions and outlook}
We have presented an in-depth study of the prototypical polar, AM Herculis, using all data obtained with \xmmn\ so far and with all instruments, supplemented by further space- and ground-based data (NuSTAR and optical photometry). The data obtained in 2015 offered a new accretion state, a reversed mode with extended phases of X-ray absorption due to an accretion curtain. Interestingly, the reversed mode is not triggered by a significantly enhanced or reduced accretion rate. 

In its lifetime AM Herculis has exhibited three distinctly different accretion modes: the normal and reversed states, and now a variant of the reversed mode with strong soft absorption. Furthermore, the anomalous slew snapshot from April 11, 2010, hints at an even richer variety of behaviors waiting to be studied. \citet{matt+00} and \cite{christian00} reported further mixed states when observed with BeppoSAX and RXTE/EUVE. 

Given the low energy resolution of the EPIC-pn we could not perform a radial velocity study of the Fe-line complex between 6.4 and 7 keV, although this was tempting given the claim of a significant velocity shift reported by \cite{girish+07}. We were, however, motivated by our study to sketch the opportunities for plasma diagnostics that will be provided in about a decade from now with ATHENA and the X-IFU as detector. We used the plasma parameters for the prime pole to normalize the output of an adapted version of the detailed MHD calculations by \citet{fischer_beuermann01} with 30 layers in the settling accretion column to simulate a 100\,ks observation of AM Herculis with ATHENA. We used a model with $B=14$\,MG, $M_{\rm WD} = 0.8$\,\msun, and specific mass flow rate $\dot{m} = 0.01$\,g\,cm$^{-2}$s$^{-1}$. The orbital velocity, thermal and velocity broadening, and the gravitational redshift were taken into account when computing phase-resolved spectra with the APEC plasma code. An inclination angle of $i=35\degr$ and a magnetic colatitude of $\beta = 65\degr$ was used for projection into the observers frame. The results for the iron line complex are shown in Fig.~\ref{f:athena} both as line plots and as an apparent trailed spectrogram to be achieved through phase-folding. We did not account for geometric fore-shortening, hence the model describes an idealized situation. Nevertheless, the prospects for detailed plasma diagnostics and radial velocity studies are breath-taking. The ionized Fe-lines originate from different layers in the accretion column which have different settling velocities. The orbital velocity is seen in the resonance line of neutral Fe, which will make AM Herculis (and similar systems) a double-lined spectroscopic binary that will allow direct measurements of the white-dwarf mass through its gravitational redshift and allow us to disentangle the physical conditions in the accretion column.

\begin{acknowledgements}
We wish to express our deep gratitude to the AAVSO amateurs that enthusiastically supported this project through dedicated, comprehensive, time-resolved photometric observations. In total 35 amateurs took part in the campaign collecting more than 16000 data points, six during the time when satellites were observing (1534 observations). We thank all the observers (listed in alphabetical order): 
John Appleyard,
Janos Bakos,
James Boardman,
Balazs Bago,
Paul Benni,
John Bortle,
Laurent Corp,
Lew Cook, 
Alicia Capetillo Blanco,
David Dowhos,
Shawn Dvorak,
Pavol Dubovsky,
Tonis Eenmae,
James Foster,
Gilles Guzman,
Tomas Gomez,
Alfredo Glez-Herrera,
Barbara Harris,
John Hall,
Miroslav Komorous,
Robert Koff, 
Damien Lemay,
Attila Madai,
Ian Miller, 
Nikolay Mishevskiy,
Kenneth Menzies,
Otmar Nickel,
Sandor Papp,
Timothy Parson,
Roger Pickard,
Gary Poyner,
George Sj\"oberg,
Seiji Tsuji,
David Vogel,
and Gary Walker. Observers from 11 countries contributed their data (in alphabetical order): Canada, Estonia, France, Germany, Great Britain, Hungary, Japan, Slovakia, Spain, Ukraine, United States of America.

This work made use of data from the NuSTAR mission, a project led by the California Institute of Technology, managed by the Jet Propulsion Laboratory, and funded by the National Aeronautics and Space Administration. We thank the \xmmn\ and NuSTAR Operations, Software and Calibration teams for support with the execution and analysis of these observations.

We thank Detlef Koester (University Kiel) for providing his model spectra of DA white dwarfs.

We thank Michael Freyberg and Frank Haberl (MPE Garching) for discussions on the calibration of EPIC-pn.

We thank C.W.~Mauche and Jan-Uwe Ness for advice on He-like triplets.

This work was supported by the German DLR under projects 50 OR 1405, 50 OR 1711, and 50 OR 1814.
\end{acknowledgements}
\bibliographystyle{aa}
\bibliography{amh2pub}

\end{document}